\documentclass[10pt,journal,compsoc]{IEEEtran}
\ifCLASSOPTIONcompsoc
  \usepackage[nocompress]{cite}
\else
  \usepackage{cite}
\fi
\ifCLASSINFOpdf
   \usepackage[pdftex]{graphicx}
   \graphicspath{{../figures/}{../jpeg/}}
   \DeclareGraphicsExtensions{.pdf,.jpeg,.png}
\else
\fi
\usepackage{algorithmic}
\usepackage{amsmath}
\usepackage{xspace,xpunctuate}
\usepackage{paralist, tabularx}
\usepackage[dvipsnames]{xcolor}
\usepackage[english]{babel}
\usepackage{hyperref}
\addto\extrasenglish{%
}

\usepackage{soul}

\newcommand{\ie}{{i.e.,}\xspace}
\newcommand{\eg}{{e.g.,}\xspace}
\newcommand{\ea}{{et~al\xperiod}\xspace}
\newcommand{\etal}{{et~al\xperiod}\xspace}

\newcommand{\cc}[1]{{(#1)}}

\newcommand{\system}{DMiner}

\newcommand{\yanna}[1]{\textcolor{black}{#1}}

\newcommand{\yannaminor}[1]{\textcolor{black}{#1}}

\newenvironment{reviseaoyu}[0]{%
    \leavevmode\color{blacks}\ignorespaces
}{}
\newenvironment{reviseyanna}[0]{%
    \leavevmode\color{black}\ignorespaces
}{}

\begin{document}


%
\title{\system: Dashboard Design Mining and Recommendation}
%
%
%
%

\author{Yanna~Lin,
        Haotian~Li,
        Aoyu~Wu,
        Yong~Wang,
        and Huamin~Qu
\IEEEcompsocitemizethanks{\IEEEcompsocthanksitem
Y. Lin, H. Li, A. Wu, H. Qu are with the Hong Kong University
of Science and Technology.\\
E-mail: \{ylindg, haotian.li, awuac\}@connect.ust.hk,  huamin@cse.ust.hk\\

\IEEEcompsocthanksitem Y. Wang is with the Singapore Management University. \\
E-mail: yongwang@smu.edu.sg
\protect\\
}
\thanks{Manuscript received April 19, 2005; revised August 26, 2015.}}

%
%

\markboth{Journal of \LaTeX\ Class Files,~Vol.~14, No.~8, August~2015}%
{Shell \MakeLowercase{\textit{et al.}}: Bare Demo of IEEEtran.cls for Computer Society Journals}
%



\IEEEtitleabstractindextext{%
\begin{abstract}
Dashboards, which comprise multiple views on a single display, help analyze and communicate multiple perspectives of data simultaneously.
However, creating effective and elegant dashboards is challenging since it requires careful and logical arrangement and coordination of multiple visualizations.
To solve the problem, we propose a data-driven approach for mining design rules from dashboards and automating dashboard organization.
Specifically, we focus on two prominent aspects of the organization: \textit{arrangement}, which describes the position, size, and layout of each view in the display space; 
and \textit{coordination}, which indicates the interaction between pairwise views.
We build a new dataset containing 854 dashboards crawled online, and 
develop feature engineering methods for describing the single views and view-wise relationships in terms of data, encoding, layout, and interactions.
Further, we identify design rules among those features and develop a recommender for dashboard design.
\yanna{We demonstrate the usefulness of \system~through an expert study and a user study.
The expert study shows that our extracted design rules are reasonable and conform to the design practice of experts.}
Moreover, a comparative user study shows that our recommender could help automate dashboard organization and reach human-level performance.
In summary, our work offers a promising starting point for design mining visualizations to build recommenders.
\end{abstract}

}


\maketitle

\IEEEdisplaynontitleabstractindextext

%
\IEEEpeerreviewmaketitle

\IEEEraisesectionheading{\section{Introduction}\label{sec:introduction}}

\maketitle

\begin{figure*}
    \centering
    \includegraphics[width=1\linewidth]{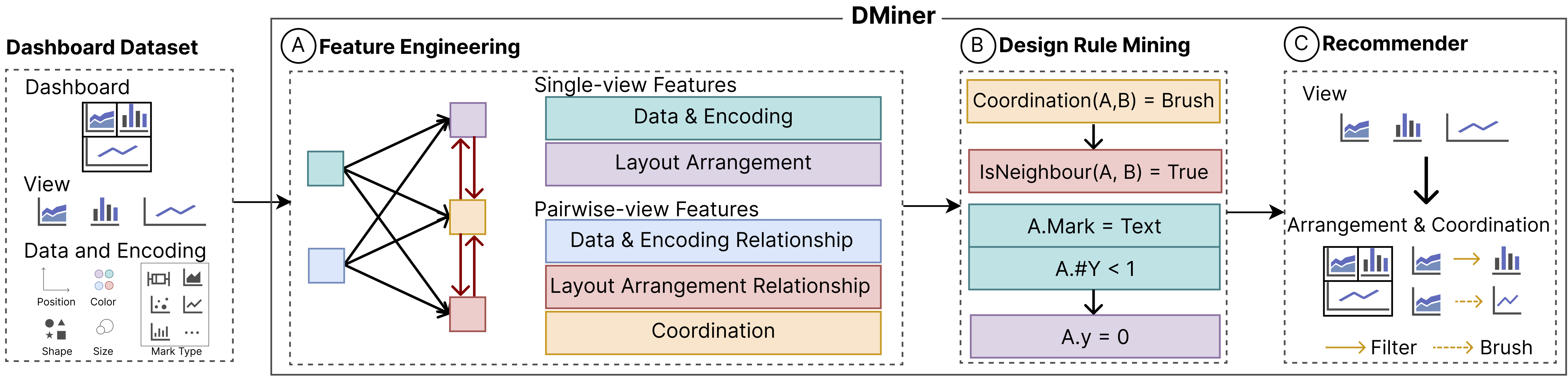}
    \caption{The workflow of \system. This paper proposes \system~as a framework for dashboard design mining \begin{reviseyanna}
    and automatic recommendation. With the dashboard dataset as the input, DMiner: \cc{A} first surveys a set of features important for dashboard design, and then extracts those features to delineate dashboard designs comprehensively. These are categorized into two types, \ie~single-view features such as data and encoding and pairwise-view features such as coordination and relative position; \cc{B} then mines design rules using decision rule approach, and further filters them; and \cc{C} finally leverages these rules for recommending dashboard arrangement and coordination.
    \end{reviseyanna}
    }
    \label{fig:teaser}
\end{figure*}


\IEEEPARstart{M}{ultiple-view} (MV) visualizations composite multiple visualizations into a single cohesive representation. 
Due to its power to support users in exploring several perspectives of data simultaneously, a large number of MV visualizations have been created and shared on the web by various domains, from biomolecular to multimedia to business.
Creating an MV visualization typically starts with selecting views of interest, followed by presenting the selected views and adding interactions between views \cite{wang2000guidelines}.
This paper refers to the presentation and interactions as layout arrangements and coordination, respectively.
Arranging and coordinating views are vital when explaining the widest range of usability problems in visualizations~\cite{camilla2010Evaluation}.
Specifically,
a proper arrangement can maximize the utility of the limited display space and improve the effectiveness and expressiveness of the information exchange, improving the usability of the system~\cite{sadana2016designing}.
Besides,
coordination among visualizations can make cross-view data relationships more apparent and reduce users' cognitive burden~\cite{chen2021nebula}. 

However,
it remains challenging to create effective MV visualizations with a proper view arrangement and coordination.
From a theoretical perspective,
existing guidelines on MV visualization designs focus on high-level recommendations,
\eg~drawing users' attention to the right \yanna{view}
and making cross-view data relationships more obvious~\cite{wang2000guidelines}.
They are insufficient in providing lay users with actionable suggestions to adjust their designs.
From a practical perspective,
while researchers have recently developed some recommenders or authoring tools to assist lay users in creating visualizations and MV visualizations,
few of them have focused on the arrangement and coordination among views.
Existing tools (\eg~Tableau~\cite{tableau}, \yanna{Power BI}~\cite{powerbi}, and MultiVision~\cite{wu2021multivision}) provide default layout templates that require manual adjustment to achieve satisfactory MV visualization designs.
This process is tedious and time-consuming,
given that the potential layouts increase exponentially with the increasing number of views.


\yanna{
We present \system, a data-driven framework for mining dashboard design and automating the layout arrangement and view coordination for MV dashboards (\autoref{fig:teaser}), thus reducing the design burden of designers.
Specifically, MV dashboards are one of the most common genres of MV visualizations \cite{sarikaya2018we}.
In this paper, we use the term MV dashboards and dashboards interchangeably to represent multiple-view dashboards.
Given that end-to-end ML-based visualization recommenders suffer from poor explainability and can confuse
end-users~\cite{saket2018beyond}, we aim to develop an explainable approach by first mining design rules from an MV dashboard dataset and further recommend appropriate layout arrangement and view coordination in MV dashboards}.

Due to the lack of MV dashboard datasets, we first crawled a large number of dashboards created by Tableau, a common dashboard authoring tool, from GitHub~\cite{github}.
We deduce a set of features that influence the arrangement and coordination of these views through reviewing prior studies, and further identify the mappings among them (\autoref{fig:teaser}~\cc{A}).
With the collected dataset, we then extract features from two perspectives: 1) the single-view features
describing each view in terms of its visual encodings (\eg color), encoded data (\eg data types), and its layout (\eg position); and 2) the pairwise-view features describing pairwise relationships between views, including the data relationship (\eg~overlapping data columns), encoding relationships (\eg~same encoding), spatial relationships (\eg~relative angle), and coordination (\eg~filter).

Building upon those features, we mine the mappings using the decision rule and filter them to distill the design rules of dashboards (\autoref{fig:teaser} \cc{B}).
Specifically, based on the data and encodings of every single view and the data relationship and encoding relationship between views, the rules infer the layout of each view and the spatial and coordination relationship between view pairs.
For instance,
one design rule in \autoref{fig:teaser}~\cc{B} shows that a text visualization (``A.Mark = Text'') without data encoding on the Y-axis (``A.\#Y $<$ 1'') tends to place at the top (``A.Y = 1'').
Using the extracted rules, we develop a recommender that recommends the optimal arrangement and coordination by ranking the obedience of design rules (\autoref{fig:teaser}~\cc{C}).

\yanna{We demonstrate the usefulness and effectiveness of \system~through an expert study and a comparative user study using the collected Tableau dashboard dataset.
We invite four experts to rate the appropriateness of the 10\% of  extracted rules, and showcase those rules with higher expert scores.
Also, we compare our recommender with the existing tool, \yanna{the crawled dashboard}, and experienced designers.}
We believe that our work can be a starting point for automating the layout arrangement and view coordination in MV dashboards.
Our main contributions are as follows:
\begin{compactitem}
    \item \system, a novel data-driven framework for dashboard design mining and recommendation.
    \item \yanna{An expert study to demonstrate the appropriateness of extracted rules and a comparative study to show the effectiveness of the recommender.}
    \item A dataset of 854 real-world dashboards with detailed information such as metadata, encodings, and coordination, which can benefit future research on dashboards.
\end{compactitem}

\section{Related Work}
\label{related_work}

Our research is related to prior studies on multiple-view visualizations, 
visualization recommendations, and design knowledge mining.

\subsection{Multiple-view Visualization}

Multiple-view visualizations (MVs) have gained extensive research interest in the visualization community.
A large body of research aims to advance the theoretical underpinnings of MVs through empirical methods.
Baldonado~\ea~\cite{wang2000guidelines} drew from a workshop to present eight guidelines for using MVs in information visualization along three dimensions: 
the selection, presentation, and interaction of views.
Roberts~\ea\cite{roberts2007state} contributed a state-of-the-art report highlighting seven fundamental research areas of MVs:
data processing, view generation, exploration techniques, coordination, tool infrastructure, human interface, usability and perception.
Sarikaya~\ea~\cite{sarikaya2018we} summarized the design space of dashboards by analyzing the intentions of a chosen corpus of 83 dashboards.
\begin{reviseyanna}
Followed by this work, Bach~\ea~\cite{bach2022dashboard} gathered 36 more dashboards and proposed a more fine-grained design space focusing on the dashboard structure, visual design, and interactivity. 
\end{reviseyanna}

Researchers have also contributed several experiments to understand the design of MVs.
For instance, Qu and Hullman~\cite{qu2017keeping} studied how visualization authors consider the importance of encoding consistency between views when designing MVs.
Langner~\ea~\cite{langner2018multiple} investigated how users interacted with MVs on large screens and concluded with design suggestions.
While those studies contributed valuable design knowledge,
they rely on large-scale experiments and require considerable human effort.
We take a different perspective by automatically extracting MV design knowledge from a larger scale corpus.

One of the key challenges in designing MVs is to relate data relationships among views~\cite{sun2021towards}.
Coordination techniques such as brushing and linking can reveal cross-view data relationships and have been widely used in visualization tools (\eg~Tableau~\cite{tableau}, Power BI~\cite{powerbi}, Jigsaw~\cite{stasko2008jigsaw}).
Despite being useful,
they require users to trigger interaction events (\eg~brushing) and pay adequate attention to subsequent changes on other views.
Thus,
researchers have advocated the use of proximity-based methods that spatially organize relationship components~\cite{sun2021sightbi}.
Inspired by those studies,
we aim to automatically recommend coordination and spatial arrangement of views to facilitate the design of MVs.
We focus on MV dashboards, which are one of the most popular genres of multiple-view visualizations~\cite{sarikaya2018we}.
Through a review of existing work, we develop a set of features that influence the arrangement and coordination of MV dashboards, extract design rules from a corpus of Tableau dashboards and further recommend appropriate arrangement and coordination of multiple views.



\subsection{\yanna{Visualization Recommendation}}
Authoring effective and elegant visualizations is a challenging task even for professionals,
since it requires the consideration of many aspects such as data insights, perceptual effectiveness, and aesthetics.
Researchers have proposed many visualization recommendation systems to assist in data analysis.
Several recommenders (\eg~\yanna{Data2Vis~\cite{dibia2019data2vis}, VizML}~\cite{hu2019vizml}, Draco~\cite{moritz2018formalizing}, and KG4Vis~\cite{li2021kg4vis}) focus on recommending visual encodings of a single visualization.

However, a single visualization is often insufficient in supporting in-depth data analysis, as it often requires progressive and iterative exploration of different data subsets~\cite{wongsuphasawat2015voyager}.
Thus, researchers have started to investigate the problem of recommending multiple visualizations.
For instance,
VizDeck~\cite{key2012vizdeck} utilizes users' preferences to select data for presentation and organize dashboards.
Voder~\cite{srinivasan2018augmenting} supports the interactive exploration of data facts associated with charts and natural language descriptions.
To further reduce manual effort,
\yanna{
Tundo \etal~\cite{tundo2020declarative} allowed users to select dashboard templates to transform the declarative definition they create into dashboards.}
MultiVision~\cite{wu2021multivision} and Dashbot~\cite{deng2022dashbot} recommend analytical dashboards given an input dataset in an end-to-end manner.
\begin{reviseyanna}
However, those systems are either template-based and do not consider the underlying data, or do not recommend layout arrangement and view coordination, which requires considerable human effort.
Our work fills this gap by developing a recommender for arranging views in an MV dashboard.
\end{reviseyanna}

Research in the field of data storytelling has studied approaches for composing multiple visualizations in a logical manner.
For instance,
researchers have proposed methods to arrange visualizations in a logical sequence to enhance storytelling~\cite{hullman2013deeper,kim2017graphscape}.
DataShot~\cite{wang2019datashot} and Calliope~\cite{shi2020calliope} group multiple visualizations into a coherent topic according to insights derived from the visualizations.
\yanna{
Different to them, we study the layout arrangement and view coordination in MV dashboards for data analysis and comprehension.}

\subsection{Design Mining Visualizations}
\begin{reviseyanna}
There are an increasing number of multiple-view visualizations created and shared by people from different domains, which has inspired many researchers to mine and extract multiple-view visualization design knowledge from them.
\end{reviseyanna}
Some researchers have invested  effort into adopting the statistical method to \yanna{mine} the visualization usage.
Al-maneea and Roberts \cite{al2019towards} collected MVs from the published papers, and answered what the most common number of views is and what the popular tiles are by statistically counting their manual-label chart type and layouts.
Beyond counting, Chen \etal\cite{chen2020composition} further leveraged some statistical methods like condition probability to perform the configuration and composition analysis, and integrated the findings into a system for exploration and recommendation.
Following this work, Shao \etal \cite{shao2021modeling} employed Bayesian probabilistic inference to analyze the effects of design factors on layouts of MVs, and discovered some insightful layout design patterns, \eg~views for exploration with a more scattered area ratio.
Though inspiring, these analyses and findings were limited to arrangements and view types, since the MVs collected in image format prevented them from accessing the underlying data and the coordination among views. 
Recently, Lu \etal \cite{lu_exploring_2020} detected the position and underlying semantics (such as the numbers 1-9) of components in infographics and explored how different components are linked.  
\yanna{Inspired by this work, we aim to mine the mappings from the characteristic of individual views and view pairs to the arrangement and coordination in an MV dashboard.}

\begin{figure*}[t!]
    \centering
    \includegraphics[width=1\linewidth]{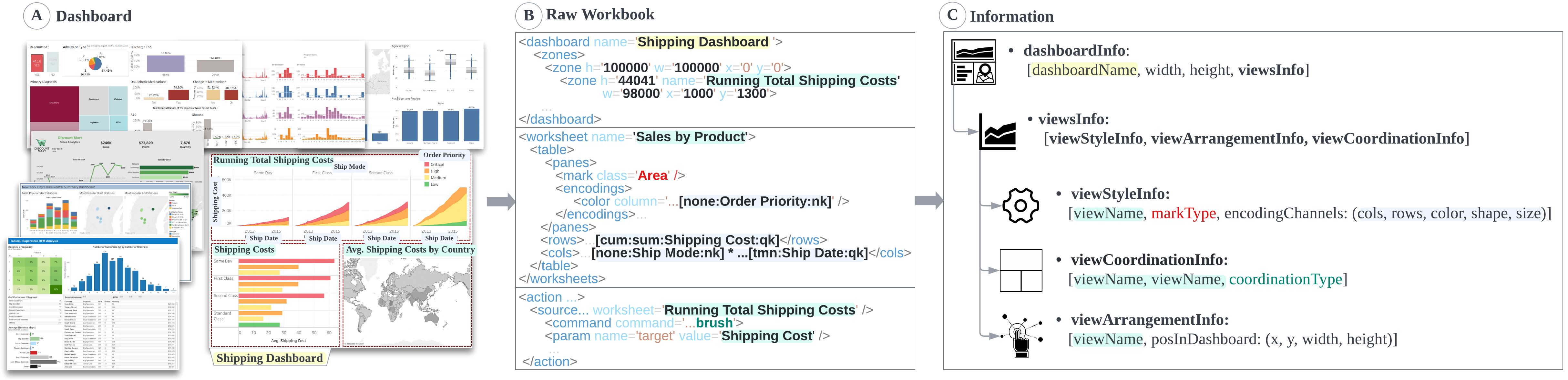}
    \caption{\yanna{Overview of the data collection pipeline: (A) We crawled a diverse set of Tableau dashboards from GitHub; (B) We automatically parsed the raw workbooks and (C) extracted information about the view arrangement and coordination.}}
    \label{fig:data_info}
\end{figure*}

\begin{figure}[t!]
    \centering
    \includegraphics[width=1\linewidth]{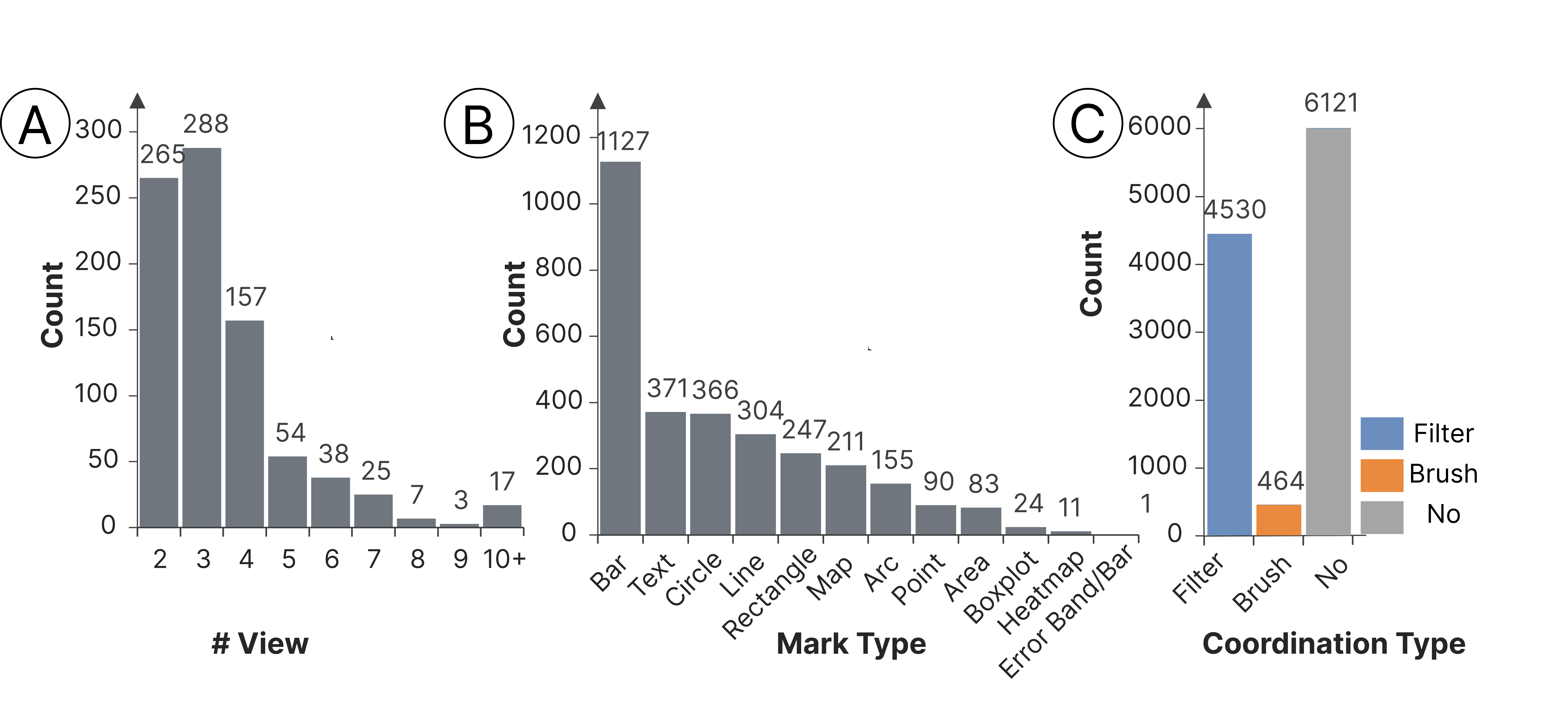}
    \caption{The basic statistics of our dashboard collection.
    \cc{A} and \cc{B} show the distribution of the number of views in each dashboard and  the distribution of the mark type of each view, respectively. \cc{C} describes the distribution of coordination types for all 854 dashboards. 
    }
    \label{fig:basic_info}
\end{figure}

\section{Dashboard Dataset}
\label{data}

Our study aims to mine design rules from existing dashboards
and further automatically recommend optimal MV dashboards.
However, there is a lack of off-the-shelf dashboard datasets that can enable automated MV dashboard design.
Specifically, existing datasets of MV dashboards~\cite{chen2020composition, shao2021modeling, al2019towards} contain only the information of individual views, such as position, size, and chart type.
They do not provide the metadata (\eg~the data operations) and the interactions between views, which are necessary information for developing visualization and dashboard recommenders~\cite{hu2019vizml,wu2021multivision}.
To address this problem, we build a new dataset from the Internet. 
The detailed procedures for constructing the dashboard datasets will be introduced as follows:

\textbf{Data Crawling.}
We first crawl Tableau dashboards from GitHub,
\ie~searching and downloading Tableau workbooks with \textit{.twbx} extensions.
There are multiple advantages to using Tableau dashboards.
First, Tableau workbooks contain all the necessary information to render the corresponding dashboards (\eg~the underlying metadata, visual encodings, and coordination among views).
Second, the information is stored in a structured XML format, which can be processed automatically at scale and can avoid the heavy manual labelling procedures required in previous studies~\cite{chen2020composition, shao2021modeling, al2019towards}.
Finally, Tableau is one of the most popular tools for creating dashboards, and its workbooks are widely shared on GitHub.
Thus, the collected dataset is diverse regarding its creators, design themes and styles, offering opportunities to mine common design rules.

\textbf{Automatic Data Processing and Cleaning.}
The datasets crawled from GitHub
suffer from noise, affecting our analysis results.
To improve the dataset quality, we first perform data cleaning by parsing and filtering the raw workbooks (.twbx).
As shown in \autoref{fig:data_info}, we parse the workbooks to derive information including the metadata (\eg~data types and data operations), encodings for each view, the coordination among views, and the arrangement of each view in a dashboard.
Subsequently, we remove workbooks without dashboards (\eg~the workbooks containing only individual views).
Dashboards using multiple data sources are also abandoned since the lack of relational schema between data sources hinders data processing.
Finally, we discard dashboards without any view coordination (\eg~cross-view interactions), since we aim to mine and recommend view coordination.
According to Tableau, there are two coordination types, namely filtering and brushing.
Specifically, \textit{filtering} removes irrelevant data objects while \textit{brushing} highlights the related visual elements and keeps the context~\cite{bartram2002filtering}.


\textbf{Manual Data Processing and Cleaning.}
Tableau workbooks offer a feature that automatically decides the mark type.
Therefore, about half of the visualizations have an ``automatic'' mark type.
\yanna{
To address this problem, we manually label the mark types according to both the definition of marks in Tableau and Vega-Lite~\cite{satyanarayan2016vega}, a widely-used grammar of visualizations.
}
A detailed list of the mark types and corresponding examples is available in {the supplementary material}.
To ensure the quality of labels, each view is labeled by two co-authors.
\yanna{The co-authors first finish the labelling individually, resulting in 99\% agreement.
Conflicts are then discussed until reaching a consensus.}
\yanna{We remove dashboards containing views with multiple mark types, \eg~the network that includes the \textit{circle} and \textit{line}.}

\textbf{Results.}
We finally collected 854 dashboards with 2990 views in total for further exploration, \yanna{and more details can be found in the supplementary material}.
Specifically, \autoref{fig:basic_info} shows the basic information of the dashboards, where \cc{A}, \cc{B}, and \cc{C} represent the distribution of the number of views of each dashboard, the mark type of each view, and the coordination among view pairs, respectively.
\autoref{fig:basic_info} \cc{A} and \cc{B} show that the distribution of the number of views aligns well with the observations from previous studies~\cite{al2019towards, chen2020composition}.
For exmaple, 2- and 3-view dashboards are the most common, and \textit{bar} is the most popular mark type.
\yanna{\autoref{fig:basic_info} \cc{C} shows that half of the view pairs possess coordination, which is almost \textit{filtering}.}

\section{DMiner}
\label{dminer}

\begin{reviseyanna}
\system{} aims to automatically extract design rules for MV dashboards from existing dashboard datasets and guide the subsequent design of MV dashboards, which is common in visualization recommenders (\eg~Draco~\cite{moritz2018formalizing}).
\autoref{fig:teaser} provides an overview of~\system{}.
We first survey previous relevant studies to identify a set of key features that indicate important considerations for MV dashboard design.
We then extract those features from our dataset and mine design rules.
Finally, we develop a recommender for MV dashboard design based on the extracted rules.
The detailed procedures of feature engineering, rule mining, and recommender will be introduced in the following subsections.
\end{reviseyanna}

\subsection{Feature Engineering}
\label{sec: feature_enginnering}
\begin{figure}[t!]
    \centering
    \includegraphics[width=1\linewidth]{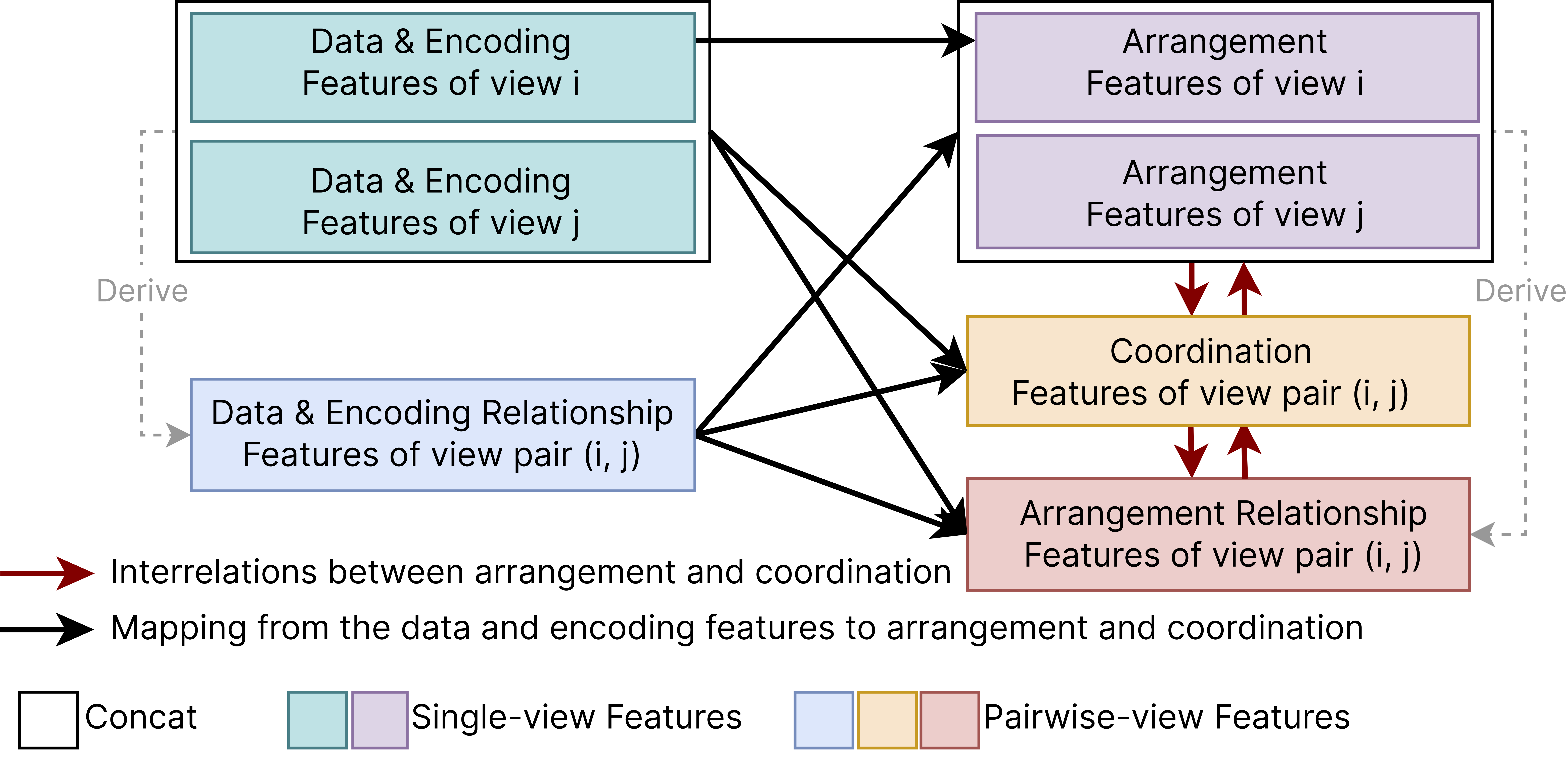}
    \caption{The targeted mappings in our design rule mining. We identified \yanna{ten mappings in black and red between features in designing dashboards, while the gray dotted lines show the derivation relationship. These black mappings focus on how to utilize data and encoding information of single views for arranging  these views in a dashboard and adding coordination among them. The red mappings between arrangement features and coordination features are used to further comprehensively describe the dashboard design.}}
    \label{fig:relationship}
\end{figure}

\autoref{fig:relationship} provides an overview of the extracted features and their mappings.
Unlike existing works that propose single-view features (\eg~the absolute position of views),
we further introduce pair-wise features (\eg~the relative position of views),
which allows us to delineate the arrangement and coordination between views in a more fine-grained manner.

\begin{figure*}
    \centering
    \includegraphics[width=1\linewidth]{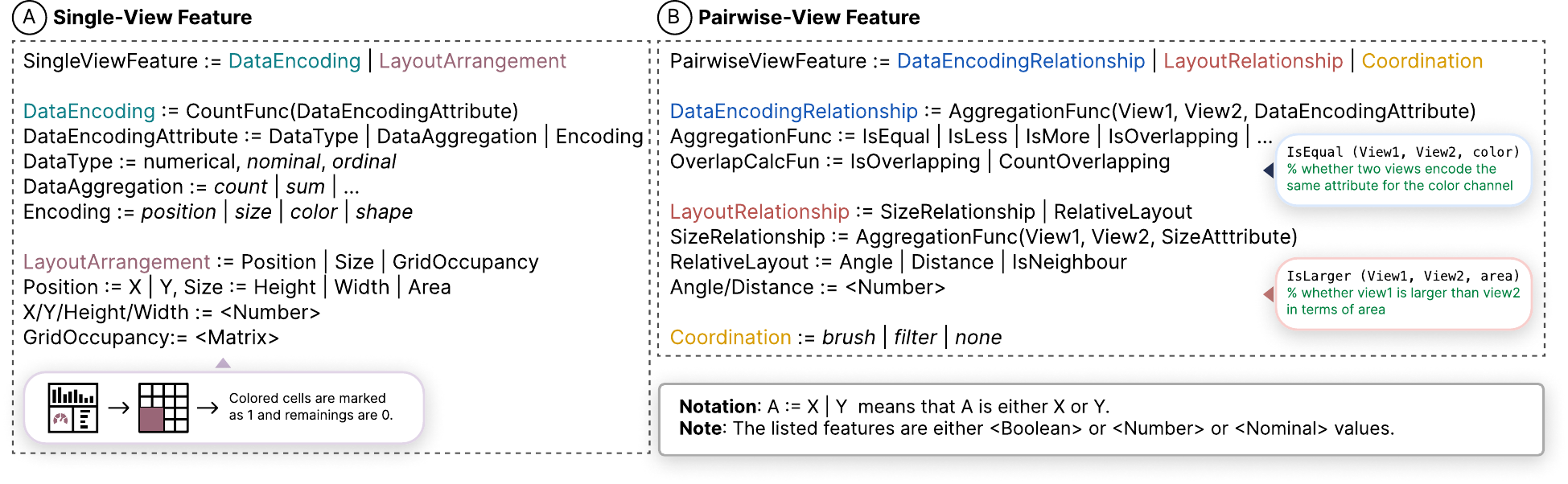}
    \caption{The formal definition and representation of extracted features. We classify features into single-view and pairwise-view features: (A) The former describes data, encoding, and layout of a single view. (B) The latter describes relationships of data, encoding, and layout between two views as well as the coordination.}
    \label{fig:features}
\end{figure*}

\begin{reviseyanna}
Our design choices for feature selection and the mappings among them are built upon prior theoretical research about dashboards and multiple visualizations~\cite{tufte1985visual, Roberts1988MV, north2000snap, chen2020composition,wu2022computableviz}.
Specifically, they have
suggested that the data and encoding features of views are crucial for arranging and coordinating the view into a cohesive dashboard.
Therefore, we choose those mappings, which are denoted as black arrows in \autoref{fig:relationship}.
\end{reviseyanna}
The bottom part of \autoref{fig:teaser}(B) shows an example mapping from \textit{data and encoding features} of a view to its \textit{arrangement features}, which describes that
a text view (i.e., \textit{A.Mark $=$ Text}) without data encoding on Y-axis (i.e., \textit{A.\#Y $<$ 1}) tends to be placed at the top (i.e., \textit{A.y $=$ 0}).
\begin{reviseyanna}
Also, prior studies have shown that the arrangement and coordination of a dashboard are interrelated~\cite{al2019towards, shao2021modeling}.
Therefore, we have further added such mappings in \system, as indicated by the red arrows in \autoref{fig:relationship}.
\end{reviseyanna}
For example, the upper part of \autoref{fig:teaser}(B) shows that two views, with one brushed by the other (\textit{coordination features}), should be put closely (\textit{arrangement relationship features}).
In summary, we have defined ten mappings among the features describing dashboard designs.
\autoref{fig:features} shows the formal definitions of all the extracted features.
They can be divided into two major groups, \ie~single-view features and pairwise-view features.
Single-view features include features about the data, encoding, and arrangement of each view in a dashboard.
Pairwise-view features are extracted to represent the relationship between two views.
Specifically, we extract the data relationship, encoding relationship, arrangement relationship, and coordination relationship between two views.
In summary, there are 33 single-view features and 41 pairwise-view features.
A detailed feature list is available in the supplementary material.


\subsubsection{Single-view Features}
Single-view features are extracted to describe each individual view in the dashboard.
These features can be roughly categorized into two types.

\textbf{Data and encoding features} can reflect what data is used in the view and how the data is visually encoded, which has been considered essential for view arrangements in a previous study~\cite{chen2020composition}.
\yanna{
Inspired by previous studies that utilize features to characterize single-view visualizations~\cite{hu2019vizml, li2021kg4vis}, we first extract data features regarding data types (\ie~numerical, nominal, and ordinal) and data operations (\eg count and sum).}
Since the numbers of data \yanna{fields} in different views are inconsistent, it is infeasible to use each field's data type and operation as features.
\yanna{Instead, as shown in \autoref{fig:features}, we define an operator, i.e., \textit{CountFunc}, to compute the frequency of different data types and data operations as features of views.}
For example, the bar chart in \autoref{fig:basic_info} \cc{A} uses one ordinal field on X-axis and one numerical field on Y-axis.
Regarding encoding information, we extract features about the mark type of the view and the usage of different encoding channels (\ie~position, size, color, shape).
Similar to data features, we also summarize the usage of each channel, for example, how many data fields are encoded using the position channel on X-axis.

\textbf{Layout arrangement features} describe how the view is placed in the dashboard. 
We propose a series of arrangement features to delineate
the
position and size information of each view.
The display area of a dashboard is evenly divided into $n \times n$ grids ($ n > 0$), and we assume that each view takes a rectangle area and does not overlap with each other.
In this paper,
$n$ is set to 4 to strike a balance between performance and efficiency in our subsequent evaluations. 
However, our mining algorithm can also work well when $n$ is other values.
As a result, besides the features delineating a view's arrangement with its original position and size, the arrangement features using the grids are also extracted.
Given that each dashboard has a different size, 
the size of all dashboards is normalized to 4*4 to enable comparisons.
That is, the width and height of each dashboard is 4, and the width and height of each view  is one of the values $[1,2,3,4]$.

\subsubsection{Pairwise-view Features} 
Previously we introduced how we extract the single-view features to characterize each individual view in the dashboard.
To further reveal the relationships across different views in the dashboard, we introduce a group of pairwise-view features.
In our study, we propose three types of pairwise-view features: 

\textbf{Data and encoding relationship features} are computed via aggregation functions on two views.
Similarly, due to the inconsistent number of fields among views, it is infeasible to calculate the statistical values of two views like correlation~\cite{hu2019vizml}. 
We propose two main types of aggregation operators shown in \autoref{fig:features}.
The first type is binary operators to compare the number of data fields, \eg~\textit{IsEqual} determines whether two views encode the same number of fields for a given encoding channel. 
\begin{reviseyanna}
The second type is set operators concerning the overlapping between data encodings, \eg~\textit{IsOverlapping} and \textit{CountOverlapping} decide whether and how many data fields encoded by two views are overlapping.

\textbf{Layout arrangement relationship features} concern the relationships of both the sizes and relative layouts of two views. For the former, we similarly apply aggregation to compare the sizes between two views, \eg~\textit{IsLarger} compares whether the size of one view is larger than the other. For the latter, we compute the relative angle and distances between two views. 
We further introduce \textit{IsNeighbour} to decide whether two views are adjacent.
\end{reviseyanna}

\textbf{Coordination features} describe the interactions between two views. 
As mentioned in \autoref{data}, there are two types of coordination in Tableau, namely brushing and filtering.
Specifically, \textit{filtering} removes irrelevant data objects while \textit{brushing} highlights the related visual elements and keeps the context~\cite{bartram2002filtering}.

\subsection{\yanna{Design Rule Mining}}
\label{sec:design_patterns}

The goal of \system{} is to find effective arrangements and coordination for the multiple views of a dashboard. 
Thus, we aim to model the mapping from data and encoding features to arrangement and coordination as well as the interrelations between arrangement and coordination within multiple dashboard views, as shown in \autoref{fig:relationship}.
Here, we adopt the decision rule approach \cite{greco2016decision} to achieve this goal. 
The reasons for choosing the decision rule approach are as follows:
1) it allows us to investigate the mutual influences among those features in a data-driven manner;
2) it embraces algorithmic explainability and is easy for humans to interpret.
Specifically, given a set of dashboard features $\mathcal{F}$ as detailed in~\autoref{fig:features},
we aim to mine rules that imply $ X \Rightarrow Y$, where $X, Y \subseteq \mathcal{F}$.
All these rules are automatically learned and extracted from the collected dashboard dataset.




\textbf{Model Selection}. As mentioned above, we have formulated the problem as mining decision rules.
To solve the problem, 
we chose one of the decision rule algorithms, RuleFit Binary Classifier~\cite{imodels2021}. 
It is efficient since it produces a set of unordered independent rules, which can be checked in parallel rather than in series~\cite{friedman2008predictive}. 
The extracted rules of RuleFit follow the structure:
$condition \rightarrow target$ with each rule's coefficient and importance.
The coefficient and importance of a rule describe how and how much the \textit{condition} contributes to the \textit{target}.
\yanna{Specifically, if the \textit{condition} is obeyed, then the corresponding \textit{target} tends to be true for positive coefficients and false for negative coefficients.}


\textbf{Feature Processing}.
Since the \textit{condition} and the \textit{target} of rules extracted by RuleFit have to be Boolean variables, we need to process the extracted features before feeding them into the model.
Specifically, we convert numerical and categorical features into Boolean variables by setting thresholds or asserting equivalence.
To handle numerical values, we set manual thresholds by splitting the interval by the average value of two-edged cut-offs (\ie~larger or smaller than the mean).
The purpose of two-edged cut-offs is to improve interpretability.
It can be challenging for humans to understand more intervals,
\eg~\textit{the distance of two views smaller than 0.5} is easier for sense-making than \textit{the distance either within $(0.0, 0.15]$ or $(0.3, 0.45]$}.
\yanna{
The final derived Boolean features are represented using binary vectors \cite{binaryvector}. }


\textbf{Training and Post-processing}. 
We train a RuleFit binary classifier for each \textit{target} feature of each mapping identified in \autoref{fig:relationship}.
In total, we train 208 models.
\yanna{
Specifically, to increase the interpretability of the rules, we reduce the condition complexity of each rule by limiting each \textit{condition} with at most 2 features~\cite{ming2018rulematrix}.}
We randomly partition dashboards into a 75\% \yanna{(640)} training set and a 25\% \yanna{(214)} test set.
The trained models achieve an average of \yanna{73\% accuracy on the training set and} 71\% accuracy  on the test set. 
Considering the huge amount of models,
we select the top 3 rules with positive coefficients and the highest importance for each model to avoid overfitting issues, resulting in 624 decision rules.


\begin{figure*}[!t]
    \centering
    \includegraphics[width=1\linewidth]{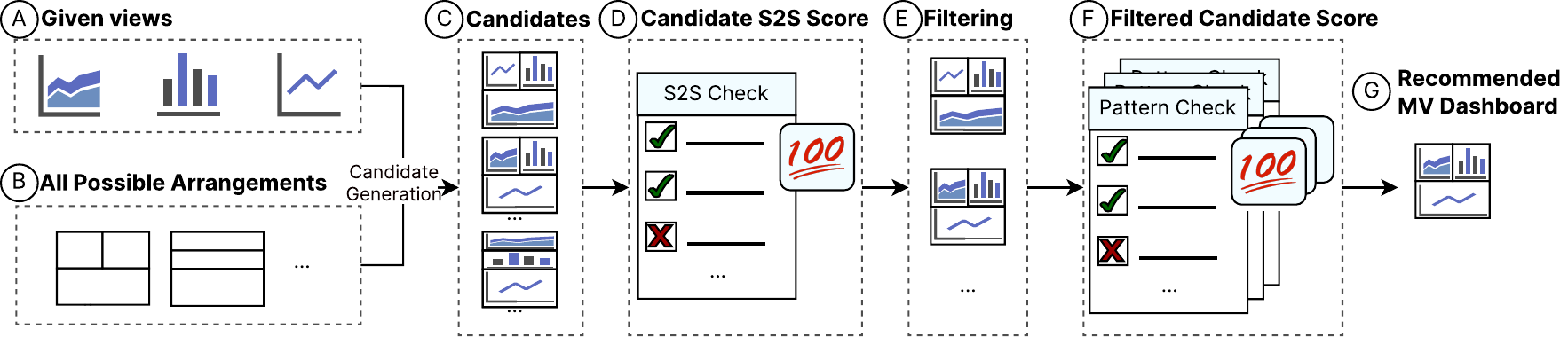}
    \caption{The recommender framework. 
    Given several views and pre-computed possible arrangements, we first generate the candidate dashboards.
    Then we apply S2S rules to prune unreasonable candidates to improve efficiency.
    The remaining candidates are further assigned scores and ranked using other rules.
    Finally, we recommend the candidate with the highest score.
    }
    \label{fig: recommendation}
\end{figure*}

\subsection{\yanna{Recommender}}
\label{sec: recommender}

Given the design rules extracted in \autoref{sec:design_patterns},
the recommender of \system{} (\autoref{fig:teaser} \cc{C}) can further automatically recommend appropriate view arrangement coordination for the views of a dashboard.
\autoref{fig: recommendation} provides an overview of the recommender pipeline.
Similar to prior optimization-based visualization recommendation approaches~\cite{wu2021ai4vis}, our recommender also
enumerates all the possible designs and  recommends the designs with the highest scores.
Specifically, it first searches all possible arrangements (\ie~positions and sizes for each view) to generate candidates.
Then, it computes the score for each candidate by checking whether it obeys or violates the extracted decision rules.
\yanna{Inspired by~\cite{moritz2018formalizing}, the cost score is a weighted sum of violated decision rules, since different rules make different contributions to the final designs}.
We use the importance of each decision rule as the weight, which measures how the rule is important to the prediction introduced in \autoref{sec:design_patterns}.

However, it is time-consuming and infeasible to calculate scores for all candidates.
The number of possible candidates is huge, since it grows exponentially with the number of views.
To reduce the computational complexity, we adopt the common strategy in previous studies on visualization recommendations (\eg SeeDB~\cite{vartak2015seedb} and QuickInsights~\cite{ding2019quickinsights}), \ie~pruning, to balance efficiency and performance.
We first consider only the rules that map \textbf{S}ingle-view data and encoding features to \textbf{S}ingle-view layout arrangement features (denoted as \textbf{S2S} rules),
since it only considers single view and has smaller complexity than pairwise-view computations.
The top 1\% recommendations with the highest average score are subsequently fed into the next step, \ie~to compute scores based on all other rules.
The percentage 1\% is decided on an empirical basis to reduce the running time to seconds, which is desirable in real-world scenarios. 
Finally, we recommend the arrangement and coordination with the highest score.

\section{Evaluation}
\label{evaluation}

\begin{reviseyanna}
To evaluate \system{}, we conducted an expert study to evaluate the appropriateness of our extracted design rules (\autoref{sec: expert_study}), as well as a user study to demonstrate the effectiveness of the recommender (\autoref{sec: user study}).

\subsection{Expert Study}
\label{sec: expert_study}

\yannaminor{
We conducted a study with four dashboard design experts to gather their quantitative and qualitative feedback about the appropriateness of the extracted design rules.
We presented the design rules extracted by our approach to the experts, and asked them to indicate the extent to which the extracted rules match their expert knowledge of dashboard designs.
To demonstrate the effectiveness of our approach, we discussed the top 5 rules with the highest appropriateness scores in detail (\autoref{tab:problematic_rules}) and summarized the feedback from the experts.
}


\subsubsection{Study Setup}


\textbf{Design Rules.}
It is time-consuming to ask our experts to evaluate all the extracted rules.
Thus, we selected Top 10\% (62) rules with the highest accuracy on the test dataset.
To increase the readability, we translated rules into natural language sentences.
For example, the rule in \autoref{fig:case example} \cc{1} was translated into \textit{``\textbf{If} View A is Text, \textbf{and} it does not have fields on Y-axis, \textbf{then} View A should be of the height of 1''}.
\yanna{Moreover}, considering that some rules may present similar conditions or semantic meanings, we further organized them into 22 \textit{processed} rules.
For example, the second and the third rule in~\autoref{tab:problematic_rules} were merged from several rules with the same \textit{if} conditions.
These 22 processed rules with the associated original rules were presented for the experts to evaluate and justify their validity.


\textbf{Participants}. We invited 4 experienced visualization or dashboard design experts (3 males, with age $31.25 \pm 2.87$). 
\textbf{Expert 1} (\textbf{E1} for short), a full professor at a local university, has engaged in visualization analytics for ten years, and has been working on MV dashboard design for the past few years.
\textbf{E2} has worked as a research scientist in a company for nine years, specializing in using dashboards for text analytics and business analysis.
\textbf{E3}, a research assistant professor, has been designing dashboards for spatial-temporal data analytics for the past seven years.
\textbf{E4}, a strategy analyst in a finance company for 3.5 years, needs to analyze trading data daily using dashboards.
None of them is color-blind.

\textbf{Procedure.}
Each expert study lasted about 1.5 hours, and we sought the experts' consent to record the entire process.
The studies were preceded by a 10-minute introduction to our work, including the goals and the corresponding features in \autoref{sec: feature_enginnering}.
In the expert studies, each expert was presented with 22 processed rules with 62 original rules. 
Also, we provided a user interface for experts to explore some supporting example dashboards for each original rule.
Experts were asked to score the appropriateness of 22 processed rules on a 7-point Likert scale from 1 (the least reasonable) to 7 (the most reasonable) and give qualitative feedback about their scores.
All the experts were encouraged to think aloud throughout the process.

\begin{figure*}[!ht]
    \centering
    \includegraphics[width=1\linewidth]{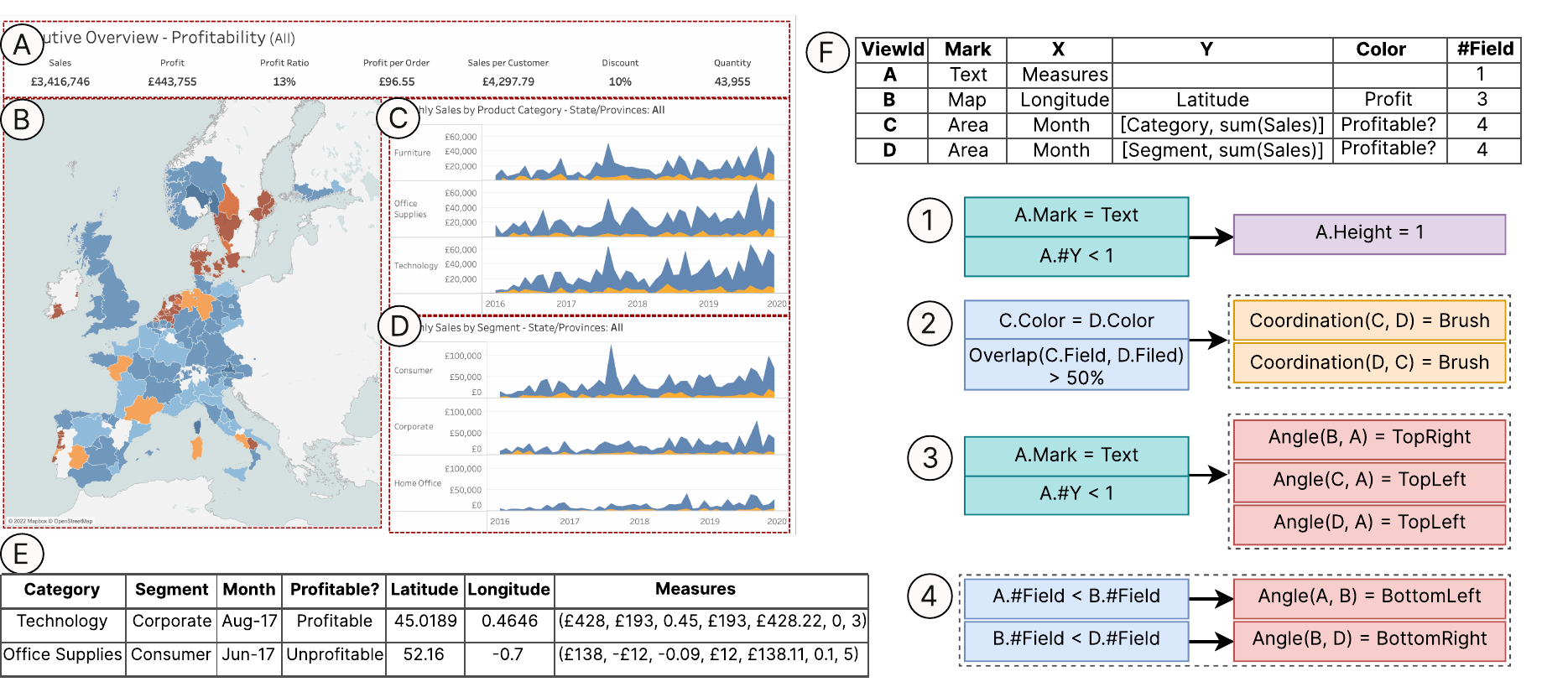}
    \caption{An example dashboard to illustrate identified design rules. The dashboard consists of four views, \cc{A} - \cc{D}. Part of the data and the information of each view in the dashboard are shown in two tables, \cc{E} and \cc{F}, respectively. In the dashboard, the application of four design rules (\cc{1} - \cc{4}) has been observed. \yanna{For example, \cc{1} represents that the height of  \textit{Text} view in \cc{A} tends to be 1, and \cc{3} describes that this \textit{Text} view in \cc{A} tends to be placed on top of other views.}}
    \label{fig:case example}
\end{figure*}

\subsubsection{Result Analysis}
Our extracted design rules were appreciated by the experts.
Considering that each processed rule represented a different number of original rules, we got a weighted average score of $4.44 \pm 1.32$, with the number of original rules as the weight.
According to the qualitative feedback from experts, our extracted design rules were reasonable and aligned well with their design knowledge, thus reducing the burden of dashboard design.
\textbf{E4} commented that our rules used \textit{low-level and more specific features}, and were therefore easier to understand and follow.
In this subsection,
we present the top five rules with the highest average scores (\autoref{tab:problematic_rules}), and summarize the feedback from the experts.
An example case is used to illustrate four of these rules, as shown in \autoref{fig:case example}.


\begin{table}[!th]
\caption{\yanna{This table shows the top 5 rules that received the highest score from the experts.}}
\label{tab:problematic_rules}
\centering
\begin{tabular}{p{0.3cm}|p{6.8cm}|p{0.7cm}}
\hline
\textbf{No.} & \textbf{Rule} & \textbf{Score} \\
\hline
\hline
1 & \textbf{If} View A is \textit{Text} (not \textit{Text} table),  \textbf{then} View A should be of the height or width of 1.               & 6.5 \\
\hline
2 & \textbf{If} View A and View B are of the same chart type, \textbf{and} they use the same fields on Y-axis,
\textbf{then} View A should be to the left or right of View B.
& 6.5 \\
\hline
3 & \textbf{If} View A and View B share more than 50\% of the same fields, \textbf{and} they use color for the same fields, 
\textbf{then} View A should brush View B or View B should brush View A.
& 6.25 \\
\hline
4 & \textbf{If} View A is \textit{Text} (not \textit{Text} table), \textbf{then} View A should be on the top right, top left, or top of other view types.  & 6 \\
\hline
5 & \textbf{If} View A has more fields than View B, \textbf{then} View A should be on the bottom right, bottom left, or bottom of View B. & 5.5\\
\hline
\end{tabular}
\end{table}

\textbf{Adjust the size of the view according to the importance of the view.}
The first rule receiving a score of 6.5 from experts is that the \textit{Text} view
(not the \textit{Text} table) 
tends to have the smallest width or height, such as \autoref{fig:case example} \cc{A} and \cc{1}.
\textbf{E2} and \textbf{E3} confirmed that ``\textit{Text view is usually used as an assistive view to describe the data analysis purpose
and does not have too many meanings.
Thus, it is often equipped with either the smallest height or the smallest width}".
\textbf{E1} mentioned that ``\textit{This reminds me that the size of the view should be related to the importance of the view, especially given the limited space available in the dashboard}".

\textbf{Configure the two views used for comparison in a similar and close manner.}
The second rule is that, if two views possess the same mark type and use the same fields on the Y-axis, they are preferred to be arranged side by side ($score = 6.5$).
All the experts inferred that these two views were for comparison, and recognized that they preferred to arrange views horizontally for comparison.
\textbf{E4} gave this rule a score of 7 and commented, ``\textit{Two views with the same chart type and Y-axis must be in the same information hierarchy in a dashboard. I would not hesitate to arrange them side by side}".
Both \textbf{E2} and \textbf{E4} said that the views for comparison should be configured ``\textit{as similar as possible, like the same width, the same height, and the same scale}".
In this way, users can compare them effectively.
While \textbf{E3} mentioned, ``\textit{In some cases, I might also arrange two comparison views vertically, if there are also other views that need to be arranged horizontally.
However, most of the time, I prefer the horizontal orientation}".
\textbf{E1} pointed out, ``\textit{This condition reminds me of the concept of `small multiples' (proposed by Tufte~\cite{tufte1985visual}), which describes the practice of arranging two views of the same visual type together to facilitate comparison}".

\textbf{Add some necessary coordination to facilitate data exploration.} 
The third rule is about coordination, \ie~if two views encode the same fields in color and use more than 50\% of the same fields, they will be coordinated by the \textit{brushing} ($score = 6.25$).
For example, as described in \autoref{fig:case example} \cc{2}, Views \cc{C} and \cc{D} have a total of 5 fields, with 4 being identical, so 80\% of the fields are shared. Moreover, both of them present the field \textit{Profitable?} using colors, so that they brush each other.
\textbf{E3} mentioned, ``\textit{This coordination is necessary, since it can help users identify the related points in another view and thus explore the data efficiently}".
\textbf{E4} pointed out, ``\textit{This is really useful if there are too many colors or objects in one view, like a scatterplot}".

\textbf{Overview first, then details-on-demand.}
The last two rules are about the relative position of views.
The first one shows that \textit{Text} view (not \textit{Text} table) tends to be on the top right, top left, or top of the other view types ($score = 5.5$).
For example, \autoref{fig:case example} \cc{A} was arranged on the top left or top right of the other three views, as described in \autoref{fig:case example} \cc{3}.
\textbf{E1} said, ``\textit{Text view providing the statistical values works as an overview and attracts the audiences' attention. Hence, the overview should be at the top of a dashboard}".
While \textbf{E2} and \textbf{E4} argued, ``\textit{it is correct to put the statistical Text view at the top, but the word cloud should not always be put at the top}". 
The last rule is that if View A encodes more fields than View B, then View A prefers to be on the bottom left, bottom right, or bottom of View B ($score = 5.5$). 
\textbf{E3} mentioned, ``\textit{the view with more fields tends to be more detailed than the view with fewer. 
According to Schneiderman's  visualization mantra `overview first, then details on demand'~\cite{shneiderman2003eyes}, the view with more fields should be at the bottom or on the right}".
\textbf{E4} agreed, ``\textit{the view with fewer fields may have more coarse information, thus having a higher hierarchy at the top}".
\textbf{E2} considered that ``\textit{the overview should be a simple view with fewer fields}".
For example, \autoref{fig:case example} \cc{4} described that View \cc{A} had the least fields, thus being arranged on top of other views.
While View \cc{D} encoded more fields than View \cc{B}, \ie~4 vs. 3, then View \cc{D} was put at the bottom right of View \cc{B}.



\textbf{Limitations and lessons.} We also received some feedback on those processed rules with low scores, which can be summarized into two main reasons, \ie~different personal experiences and shortcut learning of the algorithm.
From personal experience, experts may have different opinions on a few rules.
For example, the rule, \textit{Scatterplot should be put on the right-most}, received a score of $3.5\pm1.73$, denoted as \yannaminor{the 6th rule}.
\textbf{E1} mentioned, ``\textit{I have previously used scatterplots to show specific details, such as the distribution relationship of certain variables in two selected cities. I tend to put the detailed view on the right. Thus, I agree with this extracted rule}".
In contrast, \textbf{E3} often used T-SNE to reduce high-dimensional data to a two-dimensional scatterplot to overview the system.
Therefore, he thought ``\textit{the scatter plot should be on the left or top rather than right-most}".
Another rule with a higher score variance of 1.91 is that \textit{the view raising brushing should have a width or height of 3 or 4}, denoted as \yannaminor{the 7th rule}. 
\textbf{E4} has developed systems that use slender panels on the left or wide panels on the top. 
Thus, she thought it was reasonable for views raising \textit{brushing} to have a larger width and height.
While \textbf{E3} mentioned that ``\textit{the view proposing brushing should be a high-level view and therefore have a smaller size. 3 or 4 is too large for such a view}".
Another problem is that the decision rule approach learns some highly-correlated relationships between attributes but lack semantics for humans to interpret.
For example, 
for the rule (\textit{If View A and View B encode the same field in color, then the height of View A or View B is 3}),
\textbf{E2} commented that ``\textit{\system{} may
derive this rule from the
rules mentioned before (the 3rd rule and the 7th rule).
However, it is challenging for humans
to understand such a derivation or inference and make sense of this rule
}".

\end{reviseyanna}

\begin{figure*}
    \centering
    \includegraphics[width=1\linewidth]{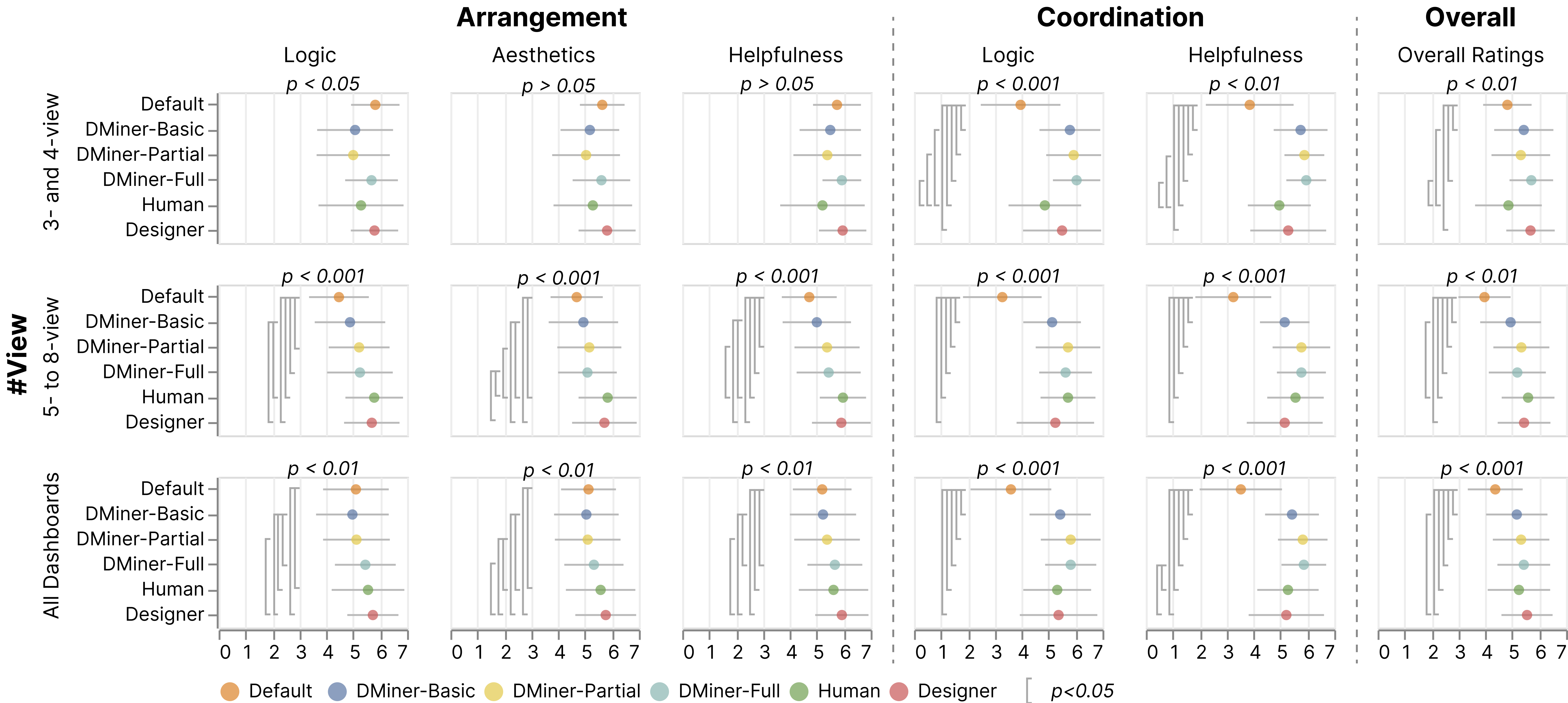}
    \caption{An overview of the study result. 
    Specifically, each column describes the score of a metric in three clusters, \ie~the cluster considering all dashboards, the cluster with 3- and 4-view dashboards, and the cluster with 5- to 8-view dashboards.
    Six metrics cover 3 perspectives, \ie~arrangement, coordination, and overall (\ie~both).
    Each sub-plot presents the score of 6 methods regarding the corresponding metric and cluster.
    The error bars are 95\% confidence intervals.
    \yanna{
    \textbf{$[$ } indicates the significant difference between a pair of methods ($p < 0.05$).}
    }
    \label{fig:result}
    \vspace{-1em}
\end{figure*}

\subsection{User Study}
\label{sec: user study}


We conducted a user study with 12 participants to \yanna{compare 6 methods} of designing MV dashboards and collect their feedback about the effectiveness of our approach for recommending appropriate MV dashboards.


\subsubsection{Study Setup}
\textbf{Participants.} 
We recruited 12 participants (9 females and 3 males, ages 22 to 27, with an average of $24.8$) \yanna{by posting online advertisements on social media platforms (\eg~email and microblogging websites)}.
They are all well-educated (7 PhD students, 3 MSc students, 1 research assistant, and 1 data analyst) and from diverse backgrounds, including \yanna{ visualization (4), data science (3), finance (2), human-computer interaction (2) and  recommendation system (1).
None of them has color blindness.
All the participants are
experienced in the data analysis according to their self-reports on a 7-point Likert scale ($\mu = 5.25, \sigma = 0.45$), where 1 indicates ``no experience'' and 7 represents ``highly experienced''.}
Each participant was compensated with \$10 after finishing the user study.

\textbf{Dashboards and Baseline Methods.}
\begin{reviseyanna}
We selected 15 dashboards from the test dataset
covering almost all the mark types (10/13) and various topics (e.g., business, COVID-19, and patient analysis). 
The number of views ranges from 3 to 8, obeying the distribution in \autoref{fig:basic_info} \cc{A}. 
To evaluate the effectiveness, we built a comparative group designed by five experienced designers
(3 females and 2 males, aged $27 \pm 2.74$).
Two are postdocs at the university, focusing on the visual analytics of computational social science and narrative storytelling, respectively.
The remaining three designers are data analysts in automobile, outlet malls, and retail companies.
They are all experienced in data analysis (with an average of 4.1 years) and dashboard design (with an average of 3 years). 
Each designer was randomly assigned three dashboards.
They were given the views of each dashboard and started their design after becoming well familiar with the view content.
Specifically, the design space of the dashboard was fixed at the common display size of 1080p (\ie~1920 x 1080)  to enable comparisons.
All designers finished the design in around 1 hour with \$15 compensation.
In this study, we leveraged 6 methods to recommend appropriate designs for 15 dashboards. The 6 methods are as follows:
\begin{compactitem}
    \item \textit{Default}: the \textit{Default} dashboard designs recommended by Tableau~\cite{tableau}. Note that Tableau does not support the recommendation of coordination. Thus, the corresponding dashboards do not have coordination;
    \item \textit{\system-Basic}: the dashboard designs recommended by \textit{\system-Basic} using design rules concerning only the single-view data and encoding features;  
    \item \textit{\system-Partial}: the dashboard designs recommended by \textit{\system-Partial} considering part of design rules, \ie~all black relationships in \autoref{fig:relationship}, and ignoring those rules on interrelations between arrangement and coordination; 
    
    \item \textit{\system-Full}: the dashboard designs recommended by \textit{\system-Full} considering all extracted design rules;
    
    \item \textit{Human}: the original dashboard designs
    we crawled online, which are designed by general \textit{Human} users; and
    \item \yanna{\textit{Designer}}: the dashboard designs created by the recruited experienced \textit{Designers}.
\end{compactitem}
\end{reviseyanna}

The above 6 methods were used to recommend dashboard designs for a given dashboard.
All the generated dashboard designs were of consistent size (i.e., 1920 x 1080 pixels), 
except for \textit{Human}, whose corresponding dashboard design was in the original size.
\yanna{For the selected 15 dashboards, it took designers around eight minutes to design a dashboard. Our three methods speeded up this process by a factor of eight. In addition, our three methods generated designs within five seconds for dashboards with fewer than six views.}
For simplicity, the resulting six dashboard designs are called \textit{a dashboard group}.
Therefore, we presented 15 dashboard groups
for participants to evaluate their design effectiveness.

\textbf{Procedure.}~The user study lasted about 1.5 hours, and we gained participants' consent for video recording the whole user study. 
During the user study, we
first briefly introduced our work.
Then,
due to the long time cost of evaluating dashboard designs, we randomly split the 15 dashboard groups into 3 clusters (each with 5 dashboard groups). 
Each participant was asked to score the arrangement and coordination of 1 cluster (\ie~5 dashboard groups).
\yanna{
For each dashboard group,
the 6 dashboard designs were ordered randomly and anonymously assigned to participants.}
Thus, each dashboard group was evaluated by 4 participants.
Before participants started to evaluate each dashboard group,
we introduced every dashboard view that needed to be arranged.
\begin{reviseyanna}
Once participants were familiar with these views, they freely explored the dashboards for around ten minutes.
After the free exploration, they further scored the dashboard designs using a 7-point Likert scale.
\end{reviseyanna}
Participants were asked to evaluate their arrangement (including logic, aesthetics, and helpfulness), coordination (including logic and helpfulness), and overall, which refers to the previous studies~\cite{smith2013data, wang2019datashot}.
Specifically, the \textit{logic} measures how well the arrangement and coordination follow human users' \textit{logical} analysis workflow~\cite{dowding2018development}.
The \textit{helpfulness} describes how well the arrangement and coordination help users analyze and explore data.
The \textit{aesthetics} evaluates how users perceive the visual appearances of the dashboard arrangement, while the \textit{overall} score measures how well the dashboard has been designed concerning both the arrangement and coordination.
All the participants were encouraged to
report why they gave the corresponding scores in a think-aloud manner.


\subsubsection{Result Analysis}
\label{subsec: quantitative_analysis}

This section introduces and discusses results from the user study, \yanna{including the ratings and the corresponding qualitative feedback.}
We received positive feedback on the recommender of \system{}.
Participants appreciated the automated dashboard designs in terms of both arrangement and coordination by our approach. 
Participant 11 (P11 for short) commented, ``\textit{I hope it can be integrated into my workflow, [...], I believe it will greatly facilitate my analysis process}".
To analyze the results, we further split all dashboards into two \textit{dashboard clusters} according to the number of views in dashboards, \ie~the dashboards with 3-4 views and the dashboards with 5-8 views.
The rationale for the split stemmed from the feedback from our participants.
They indicated that they cared more about the arrangement and coordination when the number of views is more than 4, since a well-organized dashboard can reduce their cognitive load of viewing many visualizations.
%
%
%
%
%
We performed the one-way ANOVA to compare the six methods across six metrics 
and three dashboard clusters with different view numbers (\ie~\textit{3 and 4 views}, \textit{5 to 8 views}, and \textit{all}).
LSD post-hoc tests were used if the scores obeyed the homogeneity of variance test; otherwise, Tamhane’s T2 post-hoc tests were used~\cite{lee2018proper}.
Detailed information, including the average ratings, standard deviation, and post-hoc results, are presented in \autoref{fig:result}.
\autoref{fig:result} shows that all the other methods achieved better results than \textit{Default} on all metrics, except for the 3- and 4-view dashboard arrangement.
Specifically, \textit{Default} mainly scored below 5, while the other five methods scored mostly above 5.
\begin{reviseyanna}
\end{reviseyanna}
In the following, we first introduce some interesting results regarding arrangement, coordination, and overall performance, then summarise these results, and finally discuss some recommendations that can be further improved.


\begin{figure*}[!ht]
    \centering
    \includegraphics[width=1\linewidth]{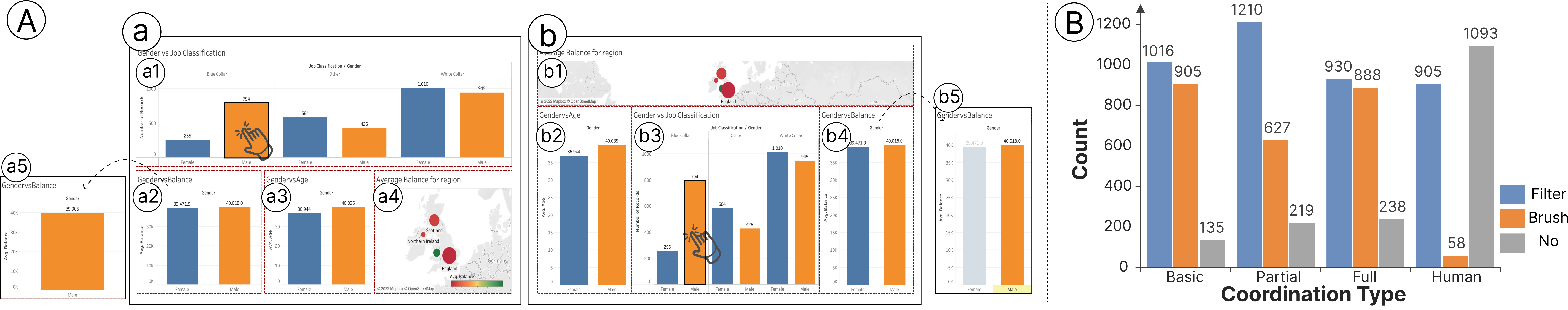}
    \caption{\cc{A} is a case identified in our user study, \yanna{and \cc{B} is the coordination type distribution.} In \cc{A}, \cc{a} and \cc{b} were the dashboards created by \textit{Human} and our method \textit{\system-Full}, respectively. Our recommended dashboard was highly rated due to outstanding arrangement logic and coordination type (see \cc{a5} and \cc{b5}).
    Participants recognized that \cc{b1} was more suitable than \cc{a1} to act as a starting point to explore the dashboard.
    For the same view pair, our recommended coordination type was \textit{brushing}, which kept the context for comparison.
    However, \textit{Human} chose \textit{filtering}, which led to a change in the display and confused the users.
    In \cc{B}, our three recommenders (\ie~\textit{\system-Basic}, \textit{\system-Partial} and \textit{\system-Full}) suggested more coordination than \textit{Human}.
    }
    \label{fig: positive_example}
\end{figure*}

\textbf{Arrangement}.
For \textbf{logic}, \autoref{fig:result} shows our method \textit{\system-Full} was not significantly different to  \yanna{\textit{Designer}} and \textit{Human} regardless of the number of views.
Specifically, it was slightly better than \textit{Human} for 3- and 4-view dashboards.
\yanna{Moreover, our method \textit{\system-Full} performed significantly better than \textit{Default} for 5- to 8-view dashboards ($p < 0.01$).}
\autoref{fig: positive_example} \cc{A} provides one example, where \cc{a} was created by \textit{Human} and \cc{b} was the recommended dashboard design by \textit{\system-Full} (with an average score of logic as 5 and 5.25, respectively).
Participants recognized that the arrangement logic of \cc{b} was better than \cc{a}, since \cc{b1} was more suitable than \cc{a1} to act as a starting point to guide the data analysis.
Furthermore, \textit{\system-Full} achieved better arrangement logic than \textit{Default} for dashboards with 5 to 8 views.
Participants acknowledged that the arrangement logic was essential for dashboards with many views, while \textit{Default} arranged the views alphabetically by view name, leading to worse logic that confused participants.
For \textbf{aesthetics}, our method \textit{\system-Full} offered a slightly better visual aesthetic design than \textit{Default}.
Participants appreciated that \textit{\system-Full} set different sizes for different views, which was better than \textit{Default} which set all views with equal width and height.
P8 mentioned that ``\textit{when there are too many views, there should be a dominant one telling me what it is attempting to express. 
The exactly-equal size confuses me, and it is challenging for me to identify what I should pay attention to first}".
Moreover, participants suggested that arrangement aesthetics was less important compared to arrangement logic for facilitating data analysis.
Thus, for  \textbf{helpfulness}, \autoref{fig:result} shows
observations similar to the logic.
\textbf{Coordination.}
\yanna{
\autoref{fig:result} shows that all our methods (\ie~\textit{\system-Full}, \textit{\system-Partial}, and \textit{\system-Basic}) offered significantly better coordination than \textit{Default} in terms of logic and helpfulness, regardless of the number of views (with all $p<0.001$).
Furthermore, \textit{\system-Partial} and \textit{\system-Full} performed similarly to \textit{Human} and \yanna{\textit{Designer}} in 5- to 8-view dashboards,
and were significantly better than \textit{Human} in 3- and 4-view dashboards (with all $p < 0.05$).}
Participants all confirmed the indispensability of the coordination among views in a dashboard.
P1 mentioned, ``\textit{Without coordination, it is challenging for me to identify the related data across views and focus the data of interest}".
As shown in \autoref{fig: positive_example} \cc{B}, compared to \textit{Human},  participants appreciated that our methods were equipped with more brushing or filtering between views for dashboards, as it provided a stronger power of exploration.
\yanna{Moreover, P5 and P11 appreciated the coordination type recommended by our approach in some cases.
They mentioned, ``\textit{\textit{Filtering} can help identify relevant information efficiently in those views with many visual elements, such as text tables with thousands of lines of text.
\textit{Brushing} works better for those views without visual clutter, since it maintains context and enables comparisons}"}.
For example, as shown in \autoref{fig: positive_example} \cc{A}, \cc{a5} was the filtered result of \cc{a2}, while \cc{b5} was the brushed result of \cc{b4}. 
It is clear to see that \cc{b5} well preserved the context before and after the interaction.
That is why our three methods gained the highest coordination score for 3- and 4-view dashboards.


\textbf{Overall.}
From \autoref{fig:result}, our method \textit{\system-Full} achieved significantly better results than \textit{Default} regardless of the number of views (with all $p<0.001$).
\yanna{Moreover, compared to \textit{Human} and \yanna{\textit{Designer}}, the average scores obtained by \textit{\system-Full} were similar without significant differences, except that \textit{\system-Full} performed significantly better than \textit{Human} for 3- and 4-view dashboards ($p < 0.01$).}
Participants mentioned that the arrangement of views in dashboards was slightly more important than the coordination for data analysis, especially for  dashboards with more views.
For 3- and 4-view dashboards, \textit{\system-Full} was significantly better than \textit{Human},
as \textit{\system-Full} can achieve similar arrangements as \textit{Human}, but with much better and more coordination. 
Regardless of the number of dashboards, our method \textit{\system-Full} achieved similar performance to \yanna{\textit{Designer}}.

\begin{figure*}[h]
    \centering
    \includegraphics[width=1\linewidth]{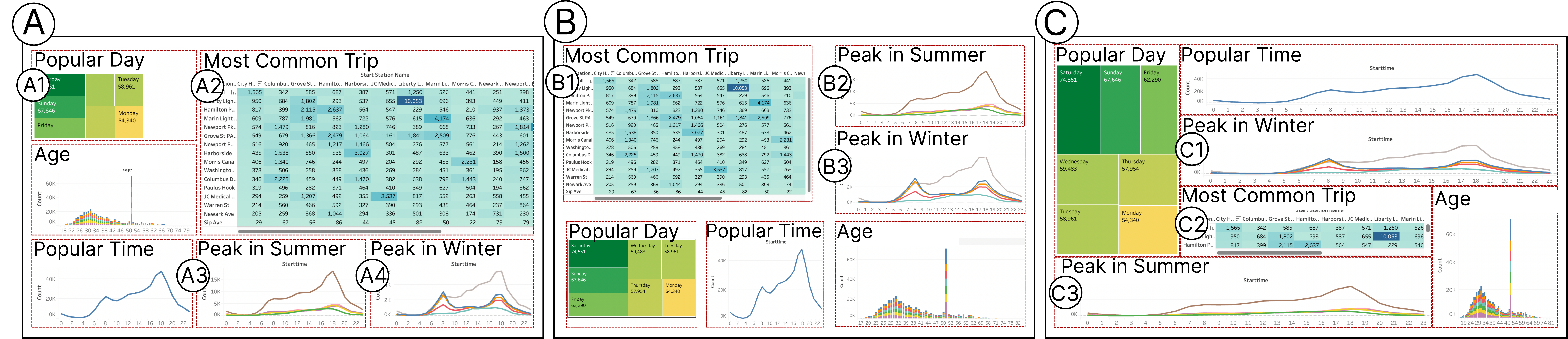}
    \caption{A case identified in our user study. 
    \yanna{\cc{A} is created by \yanna{\textit{Designer}}, \cc{B} is created by \textit{Human}, while \cc{C} is created by our method \textit{\system-Full}. 
    \cc{A} and \cc{B} received similar scores in all metrics (with a difference of less than 0.5), though they have different designs with different analysis focus.
    \textit{\system-Full}  in \cc{C} received a lower score, and our participants pointed out that the arrangements should be improved from two perspectives, \ie~the size of \cc{C2} should be equipped with more space like \cc{A2} and \cc{B1}, and \cc{C1} about ``winter" should be above \cc{C3} about ``summer".}}
    \label{fig: negative_example}
\end{figure*}

\textbf{Summary.} 
In summary,
\yanna{\textit{\system-Full} performed significantly better than \textit{Default} for dashboards with more views, and similarly to \textit{Human} and \yanna{\textit{Designer}} without significant differences.}
It was the best among our three methods, followed by \textit{\system-Partial}, and then \textit{\system-Basic}.
With an increased number of views, \textit{\system-Full} and \textit{\system-Partial} always performed better than \textit{\system-Basic}, demonstrating the importance of the pairwise-view relationship in modeling complex dashboards.
\yanna{
Moreover, no significant difference between \textit{Human} and \yanna{\textit{Designer}} in 6 metrics was observed, indicating the quality of our collected dashboard dataset.
\autoref{fig: negative_example} shows an example where \textit{Human} and \yanna{\textit{Designer}} had different designs given the same views, where \cc{A} was created by \yanna{\textit{Designer}} and \cc{B} was created by \textit{Human}.
\yanna{\textit{Designer}} regarded \cc{A1} as an overview to analyze the bike trips every day in a week,
while \textit{Human} adopted \cc{B1} as the overview to analyze the most common trips that bike riders like.
Participants recognized that they had different analysis focus, and both were designed with plausible arrangement and coordination (both with average scores larger than 5.75)}. 

\textbf{Limitations.} \system{} occasionally provides some recommendations that can be further improved.
According to participants' feedback, the underlying reasons for it stems from two perspectives, \ie~without considering the number of visual elements within each view  and the semantic meaning of data.
\yanna{Take \autoref{fig: negative_example} as an example, where the MV dashboard designed by \textit{Designer} \cc{A}, \textit{Human} \cc{B} and our method \textit{\system-Full} \cc{C} are shown.}
Compared with \autoref{fig: negative_example} \cc{A2} and \cc{B1},
the same view in the dashboard design recommended by our approach (\autoref{fig: negative_example} \yannaminor{\cc{C2}}) does not have enough space to show the heatmap.
The underlying reason for such a result is that \system{} does not explicitly consider the number of visual elements (\ie~many rectangles in the heatmap) due to the lack of underlying data tables.
Also, a few participants pointed out that \autoref{fig: negative_example} \cc{C} placed the line chart showing the \yannaminor{``Peak in Winter''} above the line chart displaying the \yannaminor{``Peak in Summer''}, which was contradictory to their expectation, \ie~the line chart for \yannaminor{``Summer"} should be positioned above that for \yannaminor{``Winter"}. Unfortunately, such semantic meanings of the data are not considered in \system{}.

\section{Discussion}
\label{sec: discussion}
In this section, we further discuss a few issues that need further clarification regarding \system{}.

\begin{reviseyanna}
\textbf{High-quality multiple-view dashboard dataset construction.}
Recent years have witnessed an increasing interest in applying AI and machine learning (ML) for visualization research~\cite{wu2021ai4vis,wang2021survey}, where high-quality datasets are fundamental to making the AI/ML-based approaches work well in practice.
However, 
little research has been
dedicated to constructing large-scale, high-quality multiple-view dashboard datasets.
In this paper, we construct a dashboard dataset by crawling GitHub repositories.
However, the designers of the dashboards published on GitHub may have different levels of visualization expertise.
Our results in \autoref{subsec: quantitative_analysis} have shown that our collected dashboard designs (i.e., \textit{Human} in \autoref{fig:result}) are not significantly different to those designed by \textit{Designer} in \autoref{fig:result}.
We believe that the quality of the dashboard designs can be further improved.
For example, appropriate dashboard designs also depend on other important factors like target users, user intention, and analysis tasks, which, however, are not included in our current dashboard dataset.
Thus, we would encourage the whole visualization community to contribute more high-quality and informative dashboard datasets.

\textbf{Human agency vs. machine automation.}
\system{} proposes a data-driven framework to mine design rules and automatically recommend dashboard designs.
Our evaluation in \autoref{subsec: quantitative_analysis} has shown that the dashboard designs recommended by \system{} (i.e., \system{}-Full) are not significantly different from those created by experienced designers. 
Recent research on visualization recommendations has tried to balance human agency and machine automation by keeping humans in the loop \cite{cao2022visguide}.
For dashboard design recommendations, it is worth further exploring how the human agency can be involved in dashboard design mining.
For instance, it will be interesting to investigate how expert knowledge can be incorporated into refining the extracted design rules for dashboards and
further improve the quality of recommended dashboards.

\textbf{\system{} for dashboards created by other software or packages.}
In this paper, \system{} extracts dashboard design rules from our collected dashboard datasets created by Tableau, one of the most popular software for creating multiple-view dashboards, and is further evaluated on only Tableau dashboards.
However, \system{}
can be extended to dashboards created using other software or packages.
For example, the source files of dashboards created by Power BI~\cite{powerbi} also specify visual coding, data operation, data types, and arrangement and coordination between views in a dashboard, which is quite similar to those of Tableau.
With a proper data parsing module, \system{} can also work for dashboards by Power BI.
However, this requires considerable research and engineering efforts to improve the interoperability,
since different visualization software and packages have different methods for specifying and rendering visualizations~\cite{satyanarayan2019critical}.
With more multiple-view dashboard datasets available in the future, we would like to  extend \system{} further to various dashboard datasets.
\end{reviseyanna}
\textbf{More factors for dashboard design.}
\system{} has considered a series of features, including data and visual encoding characteristics, for designing dashboards in terms of the arrangement and coordination of its views.
However, our evaluation result in \autoref{evaluation} has demonstrated its effectiveness.
However, there exist other factors that warrant future research.
\yanna{
As pointed out in \autoref{subsec: quantitative_analysis}, the semantic meaning of the visualized data
can also influence the arrangement and coordination of dashboard views.
For example, it is better to arrange the views showing the data of different seasons in the order of spring to winter (\autoref{subsec: quantitative_analysis}).
Also, the decision rules of \system{} focus on the effects of single features, and it will be interesting to further explore the combined effects of multiple feature types
in a single decision rule for dashboard design.}

\section{Conclusion}
\label{conclusion}
This paper proposes \system, a data-driven framework for mining dashboard design and recommending appropriate arrangement and coordination for multiple-view dashboards.
Building upon our Tableau dashboard dataset collected from GitHub,
\system{} extracts a series of relevant features, \ie~single-view and pair-view features in data and encoding, arrangement and coordination.
With these features, \system~employs a decision rule approach to mine the design rules from the collected Tableau dashboard dataset. 
Further, a recommender is proposed to recommend appropriate arrangements and coordination of dashboard views.
\begin{reviseyanna}
We conduct an expert study and a user study to evaluate the effectiveness of \system{}.
The expert study demonstrates that the extracted design rules are reasonable and align well with the design practices of experts.
The user study further confirms that the dashboard designs recommended by \system{}, considering all design rules, are not significantly different from those of experienced designers.
They are also significantly better than those using the default settings of \textit{Tableau} in both arrangement and coordination.
In summary, as the first work for automating dashboard arrangement and coordination, \system~is appreciated by visualization experts and study participants who need to design and use dashboards for data analysis. 
\end{reviseyanna}
In the future, we hope to further improve the efficiency of our recommender for dashboards with more views.
\yanna{
It is also promising to regard \system{} as a feature and integrate it into existing visualization authoring tools, such as Voyager~\cite{wongsuphasawat2015voyager}, to help designers in their daily work.}






%



\ifCLASSOPTIONcompsoc
  \section*{Acknowledgments}
\else
  \section*{Acknowledgment}
\fi
\yannaminor{We are grateful to Huan Wei, Dongyang Zhong, Zezheng Feng, Mingkai Tang, and Furui Cheng for their kind help. We also thank anonymous reviewers for their constructive comments. 
This project is partially supported by HK RGC GRF grant (16210722) and the Singapore Ministry of Education (MOE) Academic Research Fund (AcRF) Tier 2 grant (T2EP20222-0049).}

\ifCLASSOPTIONcaptionsoff
  \newpage
\fi



%

\bibliographystyle{IEEEtran}
\bibliography{main.bib}

\begin{thebibliography}{10}
\providecommand{\url}[1]{#1}
\csname url@samestyle\endcsname
\providecommand{\newblock}{\relax}
\providecommand{\bibinfo}[2]{#2}
\providecommand{\BIBentrySTDinterwordspacing}{\spaceskip=0pt\relax}
\providecommand{\BIBentryALTinterwordstretchfactor}{4}
\providecommand{\BIBentryALTinterwordspacing}{\spaceskip=\fontdimen2\font plus
\BIBentryALTinterwordstretchfactor\fontdimen3\font minus
  \fontdimen4\font\relax}
\providecommand{\BIBforeignlanguage}[2]{{%
\expandafter\ifx\csname l@#1\endcsname\relax
\typeout{** WARNING: IEEEtran.bst: No hyphenation pattern has been}%
\typeout{** loaded for the language `#1'. Using the pattern for}%
\typeout{** the default language instead.}%
\else
\language=\csname l@#1\endcsname
\fi
#2}}
\providecommand{\BIBdecl}{\relax}
\BIBdecl

\bibitem{wang2000guidelines}
M.~Q. Wang~Baldonado, A.~Woodruff, and A.~Kuchinsky, ``Guidelines for using
  multiple views in information visualization,'' in \emph{Proceedings of the
  2000 Working Conference on Advanced Visual Interfaces}, 2000, pp. 110--119.

\bibitem{camilla2010Evaluation}
C.~Forsell and J.~Johansson, ``An heuristic set for evaluation in information
  visualization,'' in \emph{Proceedings of the 2010 International Conference on
  Advanced Visual Interfaces}, 2010, p. 199–206.

\bibitem{sadana2016designing}
R.~Sadana and J.~Stasko, ``Designing multiple coordinated visualizations for
  tablets,'' in \emph{Proceedings of the Computer Graphics Forum}, vol.~35,
  no.~3, 2016, pp. 261--270.

\bibitem{chen2021nebula}
R.~Chen, X.~Shu, J.~Chen, D.~Weng, J.~Tang, S.~Fu, and Y.~Wu, ``Nebula: a
  coordinating grammar of graphics,'' \emph{IEEE Transactions on Visualization
  and Computer Graphics}, 2021 (Early access).

\bibitem{tableau}
``Tableau,''
  \url{https://help.tableau.com/current/pro/desktop/en-us/gettingstarted_overview.htm},
  [Online; accessed 2022-08-23].

\bibitem{powerbi}
``Data {Visualization} \textbar{} {Microsoft} {Power} {BI},''
  \url{https://powerbi. microsoft.com/en-us/}, [Online; accessed 2022-08-23].

\bibitem{wu2021multivision}
A.~Wu, Y.~Wang, M.~Zhou, X.~He, H.~Zhang, H.~Qu, and D.~Zhang, ``{MultiVision}:
  Designing analytical dashboards with deep learning based recommendation,''
  \emph{IEEE Transactions on Visualization and Computer Graphics}, vol.~28,
  no.~1, pp. 162--172, 2021.

\bibitem{sarikaya2018we}
A.~Sarikaya, M.~Correll, L.~Bartram, M.~Tory, and D.~Fisher, ``What do we talk
  about when we talk about dashboards?'' \emph{IEEE Transactions on
  Visualization and Computer Graphics}, vol.~25, no.~1, pp. 682--692, 2018.

\bibitem{saket2018beyond}
B.~Saket, D.~Moritz, H.~Lin, V.~Dibia, C.~Demiralp, and J.~Heer, ``Beyond
  heuristics: Learning visualization design,'' \emph{arXiv preprint
  arXiv:1807.06641}, 2018.

\bibitem{github}
``Github,'' \url{https://github.com}, [Online; accessed 2022-08-24].

\bibitem{roberts2007state}
J.~C. Roberts, ``State of the art: Coordinated \& multiple views in exploratory
  visualization,'' in \emph{Proceedings of the 5th International Conference on
  Coordinated and Multiple Views in Exploratory Visualization}, 2007, pp.
  61--71.

\bibitem{bach2022dashboard}
B.~Bach, E.~Freeman, A.~Abdul-Rahman, C.~Turkay, S.~Khan, Y.~Fan, and M.~Chen,
  ``Dashboard design patterns,'' \emph{arXiv preprint arXiv:2205.00757}, 2022.

\bibitem{qu2017keeping}
Z.~Qu and J.~Hullman, ``Keeping multiple views consistent: Constraints,
  validations, and exceptions in visualization authoring,'' \emph{IEEE
  Transactions on Visualization and Computer Graphics}, vol.~24, no.~1, pp.
  468--477, 2017.

\bibitem{langner2018multiple}
R.~Langner, U.~Kister, and R.~Dachselt, ``Multiple coordinated views at large
  displays for multiple users: Empirical findings on user behavior, movements,
  and distances,'' \emph{IEEE Transactions on Visualization and Computer
  Graphics}, vol.~25, no.~1, pp. 608--618, 2018.

\bibitem{sun2021towards}
M.~Sun, A.~Namburi, D.~Koop, J.~Zhao, T.~Li, and H.~Chung, ``Towards systematic
  design considerations for visualizing cross-view data relationships,''
  \emph{IEEE Transactions on Visualization and Computer Graphics}, 2021 (Early
  access).

\bibitem{stasko2008jigsaw}
J.~Stasko, C.~G{\"o}rg, and Z.~Liu, ``Jigsaw: supporting investigative analysis
  through interactive visualization,'' \emph{Information Visualization},
  vol.~7, no.~2, pp. 118--132, 2008.

\bibitem{sun2021sightbi}
M.~Sun, A.~R. Shaikh, H.~Alhoori, and J.~Zhao, ``{SightBi}: Exploring
  cross-view data relationships with biclusters,'' \emph{IEEE Transactions on
  Visualization and Computer Graphics}, vol.~28, no.~1, pp. 54--64, 2021.

\bibitem{dibia2019data2vis}
V.~Dibia and {\c{C}}.~Demiralp, ``{Data2Vis}: Automatic generation of data
  visualizations using sequence-to-sequence recurrent neural networks,''
  \emph{IEEE Computer Graphics and Applications}, vol.~39, no.~5, pp. 33--46,
  2019.

\bibitem{hu2019vizml}
K.~Hu, M.~A. Bakker, S.~Li, T.~Kraska, and C.~Hidalgo, ``{VizML}: A machine
  learning approach to visualization recommendation,'' in \emph{Proceedings of
  the 2019 CHI Conference on Human Factors in Computing Systems}, 2019, pp.
  1--12.

\bibitem{moritz2018formalizing}
D.~Moritz, C.~Wang, G.~L. Nelson, H.~Lin, A.~M. Smith, B.~Howe, and J.~Heer,
  ``Formalizing visualization design knowledge as constraints: Actionable and
  extensible models in draco,'' \emph{IEEE Transactions on Visualization and
  Computer Graphics}, vol.~25, no.~1, pp. 438--448, 2018.

\bibitem{li2021kg4vis}
H.~Li, Y.~Wang, S.~Zhang, Y.~Song, and H.~Qu, ``{KG4Vis}: A knowledge
  graph-based approach for visualization recommendation,'' \emph{IEEE
  Transactions on Visualization and Computer Graphics}, vol.~28, no.~1, pp.
  195--205, 2021.

\bibitem{wongsuphasawat2015voyager}
K.~Wongsuphasawat, D.~Moritz, A.~Anand, J.~Mackinlay, B.~Howe, and J.~Heer,
  ``Voyager: Exploratory analysis via faceted browsing of visualization
  recommendations,'' \emph{IEEE Transactions on Visualization and Computer
  Graphics}, vol.~22, no.~1, pp. 649--658, 2015.

\bibitem{key2012vizdeck}
A.~Key, B.~Howe, D.~Perry, and C.~Aragon, ``{VizDeck}: self-organizing
  dashboards for visual analytics,'' in \emph{Proceedings of the 2012 ACM
  SIGMOD International Conference on Management of Data}, 2012, pp. 681--684.

\bibitem{srinivasan2018augmenting}
A.~Srinivasan, S.~M. Drucker, A.~Endert, and J.~Stasko, ``Augmenting
  visualizations with interactive data facts to facilitate interpretation and
  communication,'' \emph{IEEE Transactions on Visualization and Computer
  Graphics}, vol.~25, no.~1, pp. 672--681, 2018.

\bibitem{tundo2020declarative}
A.~Tundo, C.~Castelnovo, M.~Mobilio, O.~Riganelli, and L.~Mariani,
  ``Declarative dashboard generation,'' in \emph{Proceedings of 2020 IEEE
  International Symposium on Software Reliability Engineering Workshops
  (ISSREW)}, 2020, pp. 215--218.

\bibitem{deng2022dashbot}
D.~Deng, A.~Wu, H.~Qu, and Y.~Wu, ``{DashBot}: Insight-driven dashboard
  generation based on deep reinforcement learning,'' \emph{arXiv preprint
  arXiv:2208.01232}, 2022.

\bibitem{hullman2013deeper}
J.~Hullman, S.~Drucker, N.~H. Riche, B.~Lee, D.~Fisher, and E.~Adar, ``A deeper
  understanding of sequence in narrative visualization,'' \emph{IEEE
  Transactions on Visualization and Computer Graphics}, vol.~19, no.~12, pp.
  2406--2415, 2013.

\bibitem{kim2017graphscape}
Y.~Kim, K.~Wongsuphasawat, J.~Hullman, and J.~Heer, ``{GraphScape}: A model for
  automated reasoning about visualization similarity and sequencing,'' in
  \emph{Proceedings of the 2017 CHI Conference on Human Factors in Computing
  Systems}, 2017, pp. 2628--2638.

\bibitem{wang2019datashot}
Y.~Wang, Z.~Sun, H.~Zhang, W.~Cui, K.~Xu, X.~Ma, and D.~Zhang, ``{DataShot}:
  Automatic generation of fact sheets from tabular data,'' \emph{IEEE
  Transactions on Visualization and Computer Graphics}, vol.~26, no.~1, pp.
  895--905, 2019.

\bibitem{shi2020calliope}
D.~Shi, X.~Xu, F.~Sun, Y.~Shi, and N.~Cao, ``Calliope: Automatic visual data
  story generation from a spreadsheet,'' \emph{IEEE Transactions on
  Visualization and Computer Graphics}, vol.~27, no.~2, pp. 453--463, 2020.

\bibitem{al2019towards}
H.~M. Al-maneea and J.~C. Roberts, ``Towards quantifying multiple view layouts
  in visualisation as seen from research publications,'' in \emph{Proceedings
  of the 2019 IEEE Visualization Conference (VIS)}, 2019, pp. 121--121.

\bibitem{chen2020composition}
X.~Chen, W.~Zeng, Y.~Lin, H.~M. Ai-Maneea, J.~Roberts, and R.~Chang,
  ``Composition and configuration patterns in multiple-view visualizations,''
  \emph{IEEE Transactions on Visualization and Computer Graphics}, vol.~27,
  no.~2, pp. 1514--1524, 2020.

\bibitem{shao2021modeling}
L.~Shao, Z.~Chu, X.~Chen, Y.~Lin, and W.~Zeng, ``Modeling layout design for
  multiple-view visualization via bayesian inference,'' \emph{Journal of
  Visualization}, vol.~24, no.~6, pp. 1237--1252, 2021.

\bibitem{lu_exploring_2020}
M.~Lu, C.~Wang, J.~Lanir, N.~Zhao, H.~Pfister, D.~Cohen-Or, and H.~Huang,
  ``\BIBforeignlanguage{en}{Exploring {Visual} {Information} {Flows} in
  {Infographics}},'' in \emph{\BIBforeignlanguage{en}{Proceedings of the 2020
  {CHI} {Conference} on {Human} {Factors} in {Computing} {Systems}}}, 2020, pp.
  1--12.

\bibitem{bartram2002filtering}
L.~Bartram and C.~Ware, ``Filtering and brushing with motion,''
  \emph{Information Visualization}, vol.~1, no.~1, pp. 66--79, 2002.

\bibitem{satyanarayan2016vega}
A.~Satyanarayan, D.~Moritz, K.~Wongsuphasawat, and J.~Heer, ``{Vega-Lite}: A
  grammar of interactive graphics,'' \emph{IEEE Transactions on Visualization
  and Computer Graphics}, vol.~23, no.~1, pp. 341--350, 2016.

\bibitem{tufte1985visual}
E.~R. Tufte, ``The visual display of quantitative information,'' \emph{The
  Journal for Healthcare Quality}, vol.~7, no.~3, p.~15, 1985.

\bibitem{Roberts1988MV}
J.~Roberts, ``On encouraging multiple views for visualization,'' in
  \emph{Proceedings of the 1998 IEEE Conference on Information Visualization.},
  1998, pp. 8--14.

\bibitem{north2000snap}
C.~North and B.~Shneiderman, ``Snap-together visualization: a user interface
  for coordinating visualizations via relational schemata,'' in
  \emph{Proceedings of the 2000 Working Conference on Advanced Visual
  Interfaces}, 2000, pp. 128--135.

\bibitem{wu2022computableviz}
A.~Wu, W.~Tong, H.~Li, D.~Moritz, Y.~Wang, and H.~Qu, ``{ComputableViz}:
  Mathematical operators as a formalism for visualisation processing and
  analysis,'' in \emph{Proceedings of the 2022 CHI Conference on Human Factors
  in Computing Systems}, 2022, pp. 1--15.

\bibitem{greco2016decision}
S.~Greco, B.~Matarazzo, and R.~S{\l}owi{\'n}ski, ``Decision rule approach,'' in
  \emph{Multiple criteria decision analysis}.\hskip 1em plus 0.5em minus
  0.4em\relax Springer, 2016, pp. 497--552.

\bibitem{imodels2021}
C.~Singh, K.~Nasseri, Y.~S. Tan, T.~Tang, and B.~Yu, ``{imodels: a python
  package for fitting interpretable models},'' p. 3192, 2021.

\bibitem{friedman2008predictive}
J.~H. Friedman and B.~E. Popescu, ``Predictive learning via rule ensembles,''
  \emph{The Annals of Applied Statistics}, vol.~2, no.~3, pp. 916--954, 2008.

\bibitem{binaryvector}
``Binary vector - {Wikipedia},''
  \url{https://en.wikipedia.org/wiki/Binary_vector}, [Online; accessed
  2022-08-23].

\bibitem{ming2018rulematrix}
Y.~Ming, H.~Qu, and E.~Bertini, ``{RuleMatrix}: Visualizing and understanding
  classifiers with rules,'' \emph{IEEE Transactions on Visualization and
  Computer Graphics}, vol.~25, no.~1, pp. 342--352, 2018.

\bibitem{wu2021ai4vis}
A.~Wu, Y.~Wang, X.~Shu, D.~Moritz, W.~Cui, H.~Zhang, D.~Zhang, and H.~Qu,
  ``{AI4VIS}: Survey on artificial intelligence approaches for data
  visualization,'' \emph{IEEE Transactions on Visualization and Computer
  Graphics}, 2021 (Early access).

\bibitem{vartak2015seedb}
M.~Vartak, S.~Rahman, S.~Madden, A.~Parameswaran, and N.~Polyzotis, ``{SEEDB}:
  Efficient data-driven visualization recommendations to support visual
  analytics,'' in \emph{Proceedings of the VLDB Endowment International
  Conference on Very Large Data Bases}, vol.~8, no.~13.\hskip 1em plus 0.5em
  minus 0.4em\relax NIH Public Access, 2015, p. 2182.

\bibitem{ding2019quickinsights}
R.~Ding, S.~Han, Y.~Xu, H.~Zhang, and D.~Zhang, ``{QuickInsights}: Quick and
  automatic discovery of insights from multi-dimensional data,'' in
  \emph{Proceedings of the 2019 ACM SIGMOD International Conference on
  Management of Data}, 2019, pp. 317--332.

\bibitem{shneiderman2003eyes}
B.~Shneiderman, ``The eyes have it: A task by data type taxonomy for
  information visualizations,'' in \emph{The Craft of Information
  Visualization}, 2003, pp. 364--371.

\bibitem{smith2013data}
V.~S. Smith, ``Data dashboard as evaluation and research communication tool,''
  \emph{New Directions for Evaluation}, vol. 2013, no. 140, pp. 21--45, 2013.

\bibitem{dowding2018development}
D.~Dowding and J.~A. Merrill, ``The development of heuristics for evaluation of
  dashboard visualizations,'' \emph{Applied Clinical Informatics}, vol.~9,
  no.~03, pp. 511--518, 2018.

\bibitem{lee2018proper}
S.~Lee and D.~K. Lee, ``What is the proper way to apply the multiple comparison
  test?'' \emph{Korean journal of anesthesiology}, vol.~71, no.~5, pp.
  353--360, 2018.

\bibitem{wang2021survey}
Q.~Wang, Z.~Chen, Y.~Wang, and H.~Qu, ``{A Survey on ML4VIS}: Applying
  machinelearning advances to data visualization,'' \emph{IEEE Transactions on
  Visualization and Computer Graphics}, 2021.

\bibitem{cao2022visguide}
Y.-R. Cao, X.-H. Li, J.-Y. Pan, and W.-C. Lin, ``{VisGuide}: User-oriented
  recommendations for data event extraction,'' in \emph{Proceedings of the CHI
  Conference on Human Factors in Computing Systems}, 2022, pp. 1--13.

\bibitem{satyanarayan2019critical}
A.~Satyanarayan, B.~Lee, D.~Ren, J.~Heer, J.~Stasko, J.~Thompson, M.~Brehmer,
  and Z.~Liu, ``Critical reflections on visualization authoring systems,''
  \emph{IEEE Transactions on Visualization and Computer Graphics}, vol.~26,
  no.~1, pp. 461--471, 2019.

\end{thebibliography}




%
\begin{IEEEbiography}[{\includegraphics[width=1.0in,height=1.25in,clip,keepaspectratio]{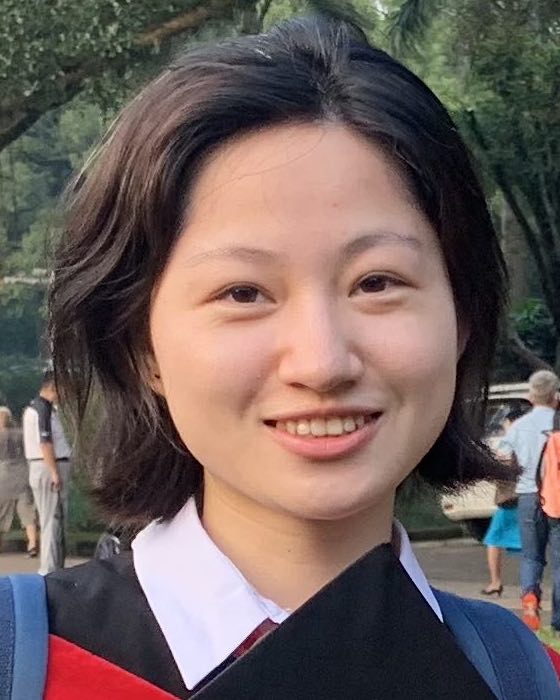}}]{Yanna Lin} is currently a PhD candidate in Computer Science and Engineering at the Hong Kong University of Science and Technology (HKUST). Her main research interests are data visualization and human-computer interaction. She received her BEng from Sun Yat-sen University. For more details, please refer to \url{https://yannahhh.github.io/}.
\end{IEEEbiography}
\begin{IEEEbiography}[{\includegraphics[width=1.0in,height=1.25in,clip,keepaspectratio]{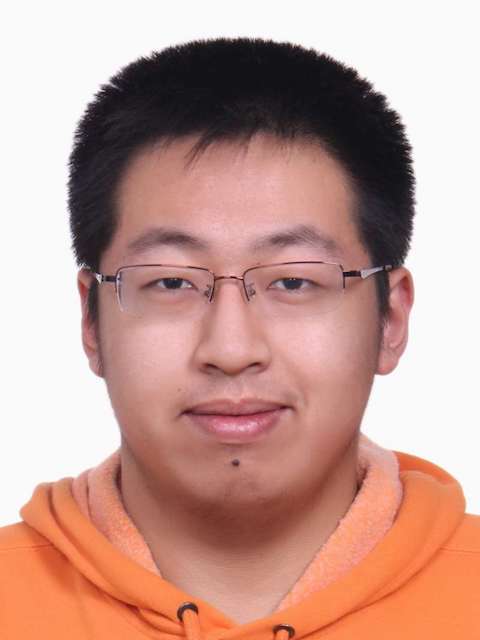}}]{Haotian Li} is currently a PhD candidate in Computer Science and Engineering at the Hong Kong University of Science and Technology (HKUST). His main research interests are data visualization, visual analytics, human-computer interaction and online education. He received his BEng in Computer Engineering from HKUST. For more details, please refer to \url{https://haotian-li.com/}.
\end{IEEEbiography}
\begin{IEEEbiography}[{\includegraphics[width=1.0in,height=1.25in,clip,keepaspectratio]{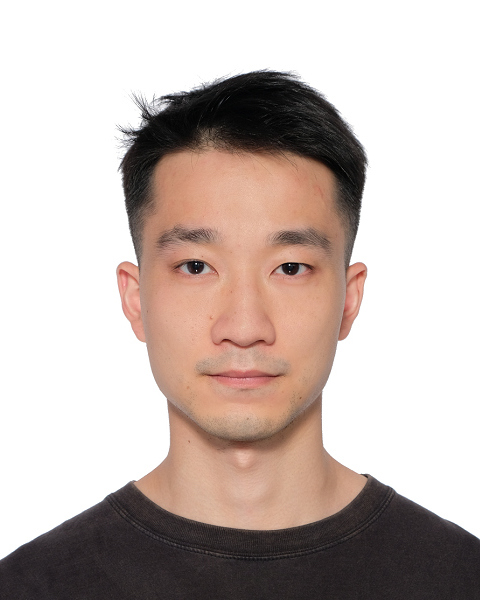}}]{Aoyu Wu} is a Ph.D. student in the Department of Computer Science and Engineering at the Hong Kong University of Science and Technology (HKUST). He received his B.E. and M.E. degrees from HKUST. His research interests include data visualization and human-computer interaction.
For more details, please refer to \url{https://wowjyu.github.io/.}
\end{IEEEbiography}
\begin{IEEEbiography}[{\includegraphics[width=1in,height=1.25in,clip,keepaspectratio]{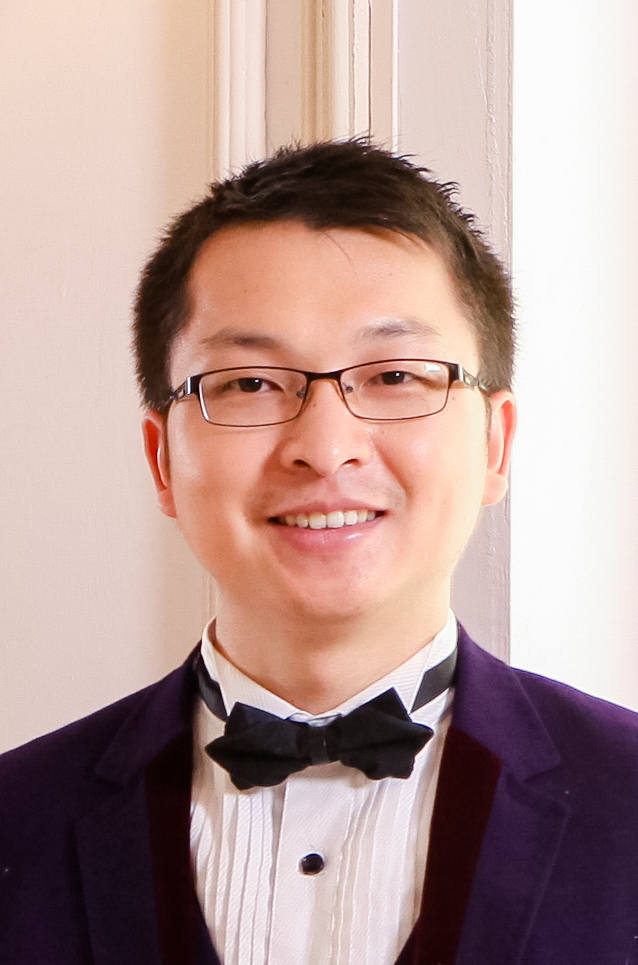}}]{Yong Wang} is currently an assistant professor in School of Computing and Information Systems at Singapore Management University. His research interests include data visualization, visual analytics and explainable machine learning.
He obtained his Ph.D. in Computer Science from Hong Kong University of Science and Technology in 2018. He received his B.E. and M.E. from Harbin Institute of Technology and Huazhong University of Science and Technology, respectively. For more details, please refer to \url{http://yong-wang.org}.
\end{IEEEbiography}
\begin{IEEEbiography}[{\includegraphics[width=1.0in,height=1.25in,clip,keepaspectratio]{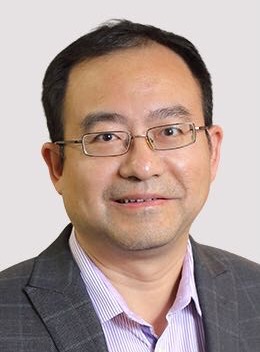}}]{Huamin Qu} is a chair professor in the Department of Computer Science and Engineering (CSE) at the Hong Kong University of Science and Technology (HKUST) and also the director of the interdisciplinary program office (IPO) of HKUST. He obtained a BS in Mathematics from Xi'an Jiaotong University, China, an MS and a PhD in Computer Science from the Stony Brook University. His main research interests are in visualization and human-computer interaction, with focuses on urban informatics, social network analysis, E-learning, text visualization, and explainable artificial intelligence (XAI). For more information, please visit \url{http://huamin.org/}.
\end{IEEEbiography}







\end{document}